\newif\ifonecolumn
\onecolumntrue
\ifonecolumn
\documentclass[onecolumn]{IEEEtran}
\else
\documentclass{IEEEtran}
\fi
\usepackage{times}
\usepackage{epsfig}
\usepackage{graphicx}
\usepackage{amssymb}
\usepackage{amsfonts}
\usepackage{amsmath}
\usepackage{booktabs}
\usepackage{amsthm}
\usepackage{pstricks,tikz}
\usepackage{multirow}
\usepackage{ifthen}
\usepackage{cite}
\usepackage{pst-node,multido}
\usepackage{enumerate}
\usepackage{paralist}

\usepackage[english]{babel}
\usepackage[utf8]{inputenc}
\usepackage{t1enc}
\newcommand{\raisedtothreefourths}[1]{\ensuremath{{#1}^{3/4}}}
\DeclareMathOperator*{\Ber}{Ber}
\newcommand{\Prob}[1]{\ensuremath{\Pr\left\{#1\right\}}}
\newcommand{\Expect}[1]{\ensuremath{{\mathbb E}\left\{#1\right\}}}
\newcommand{\sumin}{\ensuremath{\sum_{i=1}^n}}
\newcommand{\E}{\ensuremath{\mathcal{E}}}
\newcommand{\FqL}{\ensuremath{\mathbb{F}_q^L}}
\newcommand{\Fq}{\ensuremath{\mathbb{F}_q}}
\newcommand{\R}{\ensuremath{\mathbb{R}}}
\renewcommand{\S}{\ensuremath{\mathcal{S}}}
\newcommand{\A}{\ensuremath{\mathcal{A}}}
\newcommand{\s}{\ensuremath{\sigma}}

\newtheorem{theorem}{Theorem}
\newtheorem{lemma}{Lemma}
\newtheorem{definition}{Definition}

\newtheorem{corollary}{Corollary}
\newlength{\figurewidth}
\title{Secret Communication over Broadcast Erasure Channels with State-feedback}
\author{László Czap\thanks{L.~Czap is with EPFL, Switzerland. Email: laszlo.czap@epfl.ch}, Vinod M. Prabhakaran\thanks{V.~M.~Prabhakaran is with TIFR, India. Email:vinodmp@tifr.res.in}, Christina Fragouli\thanks{C.~Fragouli is with UCLA, USA and EPFL, Switzerland. Email:christina.fragouli@epfl.ch}, Suhas Diggavi\thanks{S.~Diggavi is with UCLA, USA. Email:suhas@ee.ucla.edu}%
\thanks{This paper was presented in parts at the IEEE Information Theory Workshop (ITW 2011) \cite{ITW11}, at the IEEE International Symposium on Information Theory (ISIT 2012) \cite{isit12} and at the  IEEE International Symposium on Network Coding (NetCod 2013) \cite{NetCod13}.}}

\usepackage{soul}
\newif\ifmarkchanges
\markchangestrue

\begin{document}
\maketitle

\begin{abstract}
We consider a 1-to-$K$ communication scenario, where a source transmits private messages to $K$ receivers through a broadcast erasure channel, and the  receivers feed back strictly causally and publicly their channel states after each transmission.
We explore the achievable rate region when we require that the message to each receiver remains secret - in the information theoretical sense - from all the other receivers.
We characterize the capacity of secure communication in all the cases where the capacity of the 1-to-$K$ communication scenario without the requirement of security is known.
As a special case, we characterize the secret-message capacity
of a single receiver point-to-point erasure channel with public state-feedback in the presence of a passive eavesdropper.

We find that in all cases where we have an exact characterization, we can achieve the capacity by using linear complexity two-phase schemes: in the first phase we create appropriate secret keys, and in the second phase we  use them  to encrypt each message.
We find that the amount of key we need is smaller than the size of the message,
and equal to the amount of encrypted message the potential eavesdroppers jointly collect.
Moreover, we prove that a dishonest receiver that provides deceptive feedback cannot diminish the rate experienced by the honest receivers.

We also develop a converse proof which reflects the two-phase structure of our achievability scheme. As a side result, our technique leads to a new outer bound proof for the non-secure communication problem.
\end{abstract}

\section{Introduction}

Wireless communication channels are easier to eavesdrop and harder to secure -- even towards unintentional eavesdroppers. As an example, consider a sender, Alice, who wants to send private messages to multiple (say three) receivers, Bob, Calvin and David, within her transmission radius, and assume public feedback from the receivers to Alice.   When Alice broadcasts  a message $W_1$ intended for Bob, Calvin and David should also try to overhear, as  the side information they possibly collect  can enable Alice to make her following broadcast transmissions more efficient; but then, this collected side information would allow  Calvin and David to learn parts of Bob's message.  Even worse, Calvin and David could  try to put together the parts they overheard, to extract increased information about Bob's message. 
Can we, in such a setting, keep the message for each user information theoretically secure from the other  users, even if these nodes collaborate?
Moreover, can we do so, when the users can only communicate through  shared  wireless broadcast channels? 

In this paper, we answer these questions when  communication happens through a broadcast packet erasure channel with public feedback. In particular, we assume that the receivers acknowledge through a public channel whether or not they correctly received packets; this is a natural assumption that is aligned with the operation of current wireless standards. Recent results justify the relevance of such an erasure model, e.g.~\cite{StateDep11} shows that a state dependent Gaussian channel can be viewed as a packet erasure channel. We exactly characterize the  capacity region in all the cases where the problem has been solved with no security constraints, namely, the  2-user, 3-user, symmetric $K$-user, and one-sidedly fair $K$-user \cite{Wang12,Marios2013} cases. For each such case, we present a new outer bound and a {polynomial time} achievability scheme that matches it. 

Our  achievability schemes operate in two phases:
in the first phase we efficiently generate secure keys between the source and each of the receivers, while in the second we judiciously use these keys for encryption. In both phases, we exploit channel properties to make our protocols efficient in terms of achieved rates.

In the first phase, we make use of a fundamental observation by Maurer \cite{Maurer1993}: different receivers have different looks on the transmitted signals, and we can build on these differences with the help of feedback to create secret keys \cite{Arxiv11,HeshamElGamal10}. For example, if the sender -- call her Alice -- transmits random packets through independent erasure channels with erasure probability $0.5$, there would be a good fraction of them (approximately $25\%$) that only one of two users receives, and we can transform this common randomness between Alice and the given user to a key using privacy amplification \cite{Maurer1993,Arxiv11,HeshamElGamal10,KanukurthiReyzin10}. Testbed implementations have demonstrated that a secret-key rate of several tens of Kbps is achievable by exploiting erasures in a practical wireless setting \cite{iris2013,panos2013}.

In the second phase, we use the generated keys to transmit private messages.  A naive approach is to generate secret keys {\em of the same size} as the size of the respective private messages, and then use these keys as one-time pads. However, this is too pessimistic in our case: the other users are going to receive only a fraction of the encrypted messages. 
Thus, {\em we can use  keys of smaller size than the messages}, and still be secure. To build on this observation, feedback is useful; knowing which packets a given user has successfully received, allows us to decide what to transmit next, so that we preserve secrecy from the others. 

Our schemes assume that the users provide honest feedback, but they can be extended when this is not the case. To illustrate, for the special case of $K=2$, we design a scheme that provides secrecy for an honest user even if the other user provides potentially false acknowledgments. Interestingly, we find that the same rate is achievable against dishonest users as against honest-but-curious users. We note however that, although our scheme against dishonest users is optimal in terms of achieved rates,  its security relies on the uniform distribution of the message for the dishonest party. From a practical perspective such an assumption is potentially too restrictive, which motivates us to define the new notion of distribution independent security and design a scheme that fulfills this latter, stronger security notion. We also take the opportunity to investigate the relation between different notions of security. In particular, following \cite{Vardy12} we show equivalence between our security notions and semantic security. 

To prove the optimality of our achievability schemes,  we  derive a new impossibility result for the secure $1$-to-$K$ message transmission problem, that applies for all values of $K$ and any channel parameters. Our converse proof introduces a new technique that explicitly utilizes a balance between generated and consumed keys, indicating that generating and using  keys is a natural strategy. As a side result, we provide a new proof for the known outer bound of the non-secure rate region derived in \cite{Wang12,Marios2013}.

Finally, our work also provides the secure-message capacity of a point-to-point erasure channel with public state-feedback  in the presence of a passive eavesdropper Eve. When no feedback is available, secrecy is achievable only if the legitimate receiver has a channel of larger capacity than the wiretapper \cite{Wyner75}. If public feedback is available, the work of Maurer proves that 
 a nonzero secret key generation rate is  achievable as long as the eavesdropper does not have an error free channel \cite{Maurer1993}; in this work we show we can also securely send specific messages, yet at rates lower than it is possible for key-generation. We illustrate this in \figurename~\ref{fig:itw}. 

To the best of our knowledge, this works provides the first characterization of secret-message capacity with limited public feedback for a non-trivial setup.

\section{Related work}
%============================

We distinguish between secret-key exchange, where Alice and Bob wish to agree on a common secret securely from a passive eavesdropper, Eve, and secret-message exchange, where Alice wants to send a specific message to Bob securely from Eve.

For the wiretap channel, when there is no feedback from Bob, the rates for secret key and secret-message exchange coincide; in his seminal work, Wyner derived the achievable rates of secure communication over a noisy point-to-point channel and showed that unless Eve has a worse channel than Bob, the  secure communication rate is zero \cite{Wyner75}.  These results on the wiretap channel were generalized in several directions, see e.g.~\cite{CK78,LiangMonograph}. 
The works of Maurer, Ahlswede and Csiszár have shown that in contrast, if public feedback is available, we can achieve non-zero key generation rates even if Eve has a better channel than Bob, thus establishing that public feedback can significantly increase the  achievable secure key generation rates  \cite{Maurer1993,AC93,Csiszar2004,Csiszar2008}. The wiretap channel with secure (inaccessible to Eve) feedback has been studied in \cite{Ahlswede06,Hesham,Kim}.
Our work focuses on the message exchange problem, and shows that for the case of erasure channels, public feedback  also increases the secure message exchange rate; yet, in this case, the message exchange capacity is smaller than the secret-key exchange capacity. Our results go beyond the point-to-point channel to the case of sending private messages to multiple receivers. In addition, in \cite{Maurer1993,Csiszar2008} a public channel of infinite capacity is assumed, whereas we restrict the public feedback to a short channel-state acknowledgment, which is more practical.  

Recently there has been a number of interesting works that build on physical channel characteristics (such as channel reciprocity) to derive key generation schemes \cite{JMDF11,JDFPA10}. Our work focuses on erasure channels and on the message sending problem  \cite{ITW11,isit12}.
Secret-message exchange has also been studied over networks in the case where there are no channel errors \cite{Yeung2005a}; in our setup, the channel variability is what makes secrecy possible. 

Finally, as mentioned in the introduction, the use of feedback and broadcast for private message transmission without security requirements has been studied in \cite{Leandros09,MaddahAliTse10,Marios2013}.

\section{Definitions and background}

A sender, Alice, wants to send private messages to a set of $K$ receivers: she wants to send message $W_j$ to  receiver $j$, so that, no other receiver learns $W_j$, even if all other receivers collude. 

\subsection{Communication model}
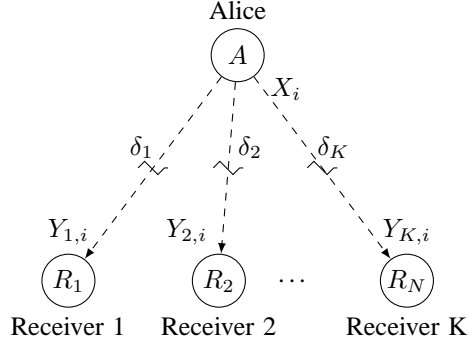
\begin{figure}
\centering
\newsavebox\Csize
\sbox{\Csize}{\tikz \node[circle,inner sep=1pt]{$R_K$};}
\begin{tikzpicture}[>=latex]
\usetikzlibrary{snakes}
\node (A) [circle, draw,minimum width=\wd\Csize,label=Alice,inner sep = 0pt,label=-30:$X_{i}$] at (-0.75,1) {$A$};
\node (B) [circle, draw,minimum width=\wd\Csize,label=below:Receiver~1,inner sep = 0pt,label=above:$Y_{1,i}$] at (-3,-2) {$R_1$};
\node (C) [circle, draw,minimum width=\wd\Csize,label=below:Receiver~2,inner sep = 0pt,label=95:$Y_{2,i}$] at (-1,-2) {$R_2$};
\node at (0,-2){\dots};
\node (D) [circle, draw,minimum width=\wd\Csize,label=below:Receiver~K,inner sep = 0pt,label=above:$Y_{K,i}$] at (1.5,-2) {$R_N$};
\draw[dashed,->] (A)--(B) node (AB)[midway]{};
\draw[dashed,->] (A)--(C) node (AC)[midway]{};
\draw[dashed,->] (A)--(D) node (AD)[midway]{};
\draw[snake=zigzag] (AB) ++(-.2,0) -- ++(.4,0) node [midway,above,very near start]{$\delta_1$};
\draw[snake=zigzag] (AC) ++(-.2,0) -- ++(.4,0) node [midway,above,very near end,xshift=4]{$\delta_2$};
\draw[snake=zigzag] (AD) ++(-.2,0) -- ++(.4,0) node [midway,above,very near end]{$\delta_K$};
\end{tikzpicture}
\caption{1-to-$K$ broadcast erasure channel}
\label{fig:model}
\end{figure}

Communication takes place over a \mbox{1-to-$K$} broadcast erasure channel, with input at Alice and an output at each of the $K$ receivers. We illustrate the setting in \figurename~\ref{fig:model}. We also summarize our frequently used notations in Table~\ref{tab:notations}. The channel input alphabet consists of all possible vectors of length $L$  over a finite field $\Fq$. For convenience, we usually call such a vector a \emph{packet}. Throughout the paper we express entropy and rate in terms of packets. This enables  us to omit the constant factor $L\log q$. 

\begin{table}
\caption{Summary of notation}
\label{tab:notations}
\center
      \begin{tabular}{cl}
	\toprule
	$X_i,Y_{j,i}	$		&The $i$th input and outputs of the channel \\
	$S_i,S_i^*$			&The actual and the acknowledged state of the channel in the $i$th transmission\\
	$\delta_j$			&Erasure probability for receiver~$j$\\
	$W_j$				&Private message for receiver~$j$\\
	$P_{W_i},P_{W_i,W_j}$		&Distribution and joint distribution of messages $W_i$, $W_j$\\
	$\s_1, \s_2$			&Acknowledging strategy of a dishonest user\\	
	$N_j$				&Size of $W_j$ expressed in number of packets\\
	$R_j$				&Secret message rate for receiver~j\\
	$\Theta_A,\Theta_j$	&Private randomness of Alice and of receiver~$j$ \\		
	\bottomrule
      \end{tabular}      
\end{table}

The broadcast channel is made up of $K$ independent\footnote{We assume independence for simplicity, but as long as the statistical behavior of the channel is known, our results can be easily generalized.} component erasure channels with packet erasure probabilities $\delta_1,\delta_2,\dots,\delta_K$. Below we define the channel formally (see Eq.~\eqref{eq:ch_def1}-\eqref{eq:ch_def2}). We assume that the receivers send public acknowledgments after each transmission stating whether or not they received the transmission correctly. By \emph{public} we mean that the acknowledgments are available not only for Alice but for all other receivers as well. We assume that some authentication method prevents the receivers from forging each other's acknowledgments.  Also, receivers learn each other's acknowledgments causally, after they have revealed their own. 

Let $S_i\in 2^{\{1,2,\dots,K\}}$ denote the state of the channel in the $i$th transmission. $S_i$ collects the indices of receivers with correct reception. In the first place, we assume that acknowledgments are honest. We relax this assumption for the special case of $K=2$.
%Further, $S_i^*$ denotes the state based on the acknowledgments sent by the receivers. If all receivers report honestly, then \mbox{$\forall\  i: S_i=S_i^*$}.

We denote by $X_i$ the $i$th transmission over the channel, and by $Y_i=(Y_{1,i},Y_{2,i},\dots,Y_{K,i})$ the corresponding outputs observed by the receivers. We use $X^n$ to denote the vector $(X_1,X_2,\dots,X_n)$. We use a similar notation for other vectors as well.
Formally, the channel behavior is defined as
\begin{align}
\Prob{Y_i|X_i}&=\prod_{j=1}^K\Prob{Y_{j,i}|X_i} \label{eq:ch_def1}\\
\forall\  j\in\{1,2,\dots,K\}:\Pr\{Y_{j,i}|X_i\}&=\begin{cases}
1-\delta_j, &Y_{j,i}=X_i\\
\delta_j, &Y_{j,i}=\perp,\label{eq:ch_def2}
\end{cases}
\end{align}
where $\perp$ is the symbol of erasure.

We assume that all participants can generate private randomness at unlimited rate. We denote the private randomness of the sender $\Theta_A$, while $\Theta_j$ is the private randomness of user~$j$. Variables  $\Theta_A$ and every $\Theta_j$ are independent from all other randomness in the system.

\subsection{Reliability and security -- honest-but-curious users}
An $(n,\epsilon,N_1,N_2,\dots,N_K)$ scheme sends message $W_j$ which consist of $N_j$ packets of length $L$ to receiver $j$ using $n$  transmissions from Alice with error probability smaller than $\epsilon$. We denote $W=(W_1,W_2,\dots,W_K)$ the set of all messages.

\begin{definition}
\label{def:scheme}
An $(n,\epsilon,N_1,N_2,\dots,N_K)$ {\em scheme} for the \mbox{1-to-K} broadcast channel consists of the following components: (a) message alphabets ${\mathcal
W}_j={\mathbb F}_q^{LN_j}$, $j=1,2,\dots,K$, (b)
encoding maps $f_i(.)$, $i=1,2,\ldots,n$, and (c) decoding maps $\phi_j(.)$, $j=1,2,\dots,K$, such that the inputs to the channel are
\begin{align}
X_i = f_i(W,\Theta_A,S^{i-1}),\quad i=1,2,\ldots, n,
\label{eq:ch_inputs}
\end{align}
where messages $W_j\in{\mathcal W}_j$ are arbitrary messages in their respective alphabets and $\Theta_A$ is the private randomness Alice has access to. Further, it provides decodability for each receiver, that is
\begin{align}
\forall\  1\leq j\leq K: \Pr\{\phi_j(Y_j^nS^n)\neq W_j\} &< \epsilon \label{eq:decodability} 
\end{align}
is satisfied.
\end{definition}

\begin{definition}
\label{def:rate_region}
The capacity region of the \mbox{1-to-$K$} broadcast erasure channel \mbox{$\mathcal{R}^K\subset \R_+^K$} is the set of rate tuples for which for every \mbox{$\epsilon >0$} there exists an $(n,\epsilon,N_1,N_2,\dots,N_K)$ scheme that satisfies 
\begin{align}
\forall\  1\leq j\leq K: R_j-\epsilon<\frac{1}{n}N_j \label{eq:rates}
\end{align}
for all $1\leq j\leq K$.
\end{definition}
The following definition extends Definition~\ref{def:scheme} with a security requirement.
\begin{definition}
\label{def:secrecy}
An $(n,\epsilon,N_1,N_2,\dots,N_K)$ scheme is {\em secure against honest-but-curious users} if  in addition to \eqref{eq:ch_inputs}-\eqref{eq:decodability} the following also holds for all $1\leq j\leq K$:
\begin{align}
\max_{P_W} I(W_j;Y_{-j}^nS^n\Theta_{-j}) < \epsilon \label{eq:secrecy},
\end{align}
where the maximum is taken over all possible joint message distributions and $Y_{-j}^n$ is a shorthand for $Y_{1}^n,\dots,Y_{j-1}^n,Y_{j+1}^n,\dots,Y_K^n$. Similarly, $\Theta_{-j}$ is $\Theta_1,\dots,\Theta_{j-1},\Theta_{j+1},\dots,\Theta_{K}$.
\end{definition}

\begin{definition}
\label{def:rate_region_honest}
The secret-message capacity region of the \mbox{1-to-$K$} broadcast erasure channel \mbox{$\mathcal{R}^K_H\subset \R_+^K$} is the set of rate tuples for which for every \mbox{$\epsilon >0$} there exists an $(n,\epsilon,N_1,N_2,\dots,N_K)$ scheme that is secure against honest-but-curious users and satisfies
\begin{align}
\forall\  1\leq j\leq K: R_j-\epsilon<\frac{1}{n}N_j\label{eq:rates_honest}.
\end{align}
\end{definition}
Following \cite{Wang12} we distinguish two special cases.

\begin{definition}
We call the channel \emph{symmetric} if the erasure probabilities are all the same: $\delta_i=\delta_j, \forall\  1\leq i,j\leq K$.
\end{definition}

\begin{definition}
We call a rate vector one-sidedly fair if $\delta_i \geq \delta_j$ for $i\neq j$ implies
\begin{align}
R_i\delta_i \geq R_j\delta_j.
\end{align}
\end{definition}

\subsection{Dishonest users}

For the special case of $K=2$ we introduce further, stronger notions of security. In particular, we aim to provide secrecy against receivers who might acknowledge dishonestly. For convenience, we call the receivers Bob and Calvin. A \emph{dishonest user} can produce dishonest acknowledgments as a (potentially randomized) function of all the information he has access to when producing each acknowledgment (this includes all the packets and the pattern of erasures he received up to and including the current packet he is acknowledging and the acknowledgments sent by the other user over the public channel up to the previous packet).  In the following $\mathcal{S}_1$ and $\mathcal{S}_2$ denote the set of all possible acknowledging strategies of Bob and Calvin respectively and $\s_1 \in \mathcal{S}_1$ and $\s_2 \in \mathcal{S}_2$ denote their elements.

We do not provide any guarantees for a dishonest user, hence at most one of the two receivers can be dishonest, otherwise the problem would not be meaningful. Of course, the sender, Alice is not aware of which user is dishonest, otherwise she could simply ignore the dishonest party.

Our security definition for the case of honest-but-curious users does not depend on the joint distribution of the messages. 
In contrast, we define security against a dishonest user under three different assumptions on the joint message distribution that correspond to different levels of security. First, we assume that messages are independent and uniformly distributed (see Definitions~\ref{def:security_dishonest_uniform}-\ref{def:rate_region_dishonest_uniform}). This models the case when messages are properly source-coded and the users have no control on them. Second, we relax any assumption on the message distribution (see Definitions~\ref{def:security_dis}-\ref{def:rate_region_dis}). This model even allows that the dishonest user selects arbitrarily the joint distribution of the messages. Third, we assume that messages are independent and the message of the  dishonest user is uniformly distributed, but we do not make any assumption on the message distribution of the honest user (see Definitions~\ref{def:security_dishonest}-\ref{def:rate_region_dishonest}). In this model, the dishonest user might choose the distribution of only the other user's message. According to another interpretation, in this case, the dishonest user might have side information about the message distribution of the honest user. Security in this last model also ensures resistance against a chosen-plaintext attack.

We denote by $S^*_i$ the $i$th channel state based on the acknowledgments from Bob and Calvin. If there is a dishonest user, then potentially $S_i\neq S_i^*$, thus Alice and the honest party do not have access to the true channel states. We need to modify Definition~\ref{def:scheme} for $K=2$ accordingly: 
\begin{definition}
\label{def:dishonest_scheme}
An $(n,\epsilon,N_1,N_2)$ {\em scheme} for the two user message transmission
problem consists of the following components: (a) message alphabets ${\mathcal
W}_1={\mathbb F}_q^{LN_1}$ and ${\mathcal W}_2={\mathbb F}_q^{LN_2}$, (b)
encoding maps $f_i(.)$, $i=1,2,\ldots,n$, and (c) decoding maps $\phi_1(.)$ and
$\phi_2(.)$, such that if the inputs to the channel are
\begin{align}
X_i = f_i(W_1,W_2,\Theta_A,S^{\ast i-1}),\quad i=1,2,\ldots, n,
\label{eq:def1_1}
\end{align}
where $W_1\in{\mathcal W}_1$ and $W_2\in{\mathcal W}_2$ are arbitrary
messages in their respective alphabets and
$\Theta_A$ is the private randomness Alice has access to. Then,
provided the receivers acknowledge honestly,
\begin{align}
\Pr\{\phi_1(Y_1^nS^{*n}) \neq W_1\} &< \epsilon,\text{ and} \label{eq:def1_2} \\
\Pr\{\phi_2(Y_2^nS^{*n}) \neq W_2\} &< \epsilon. \label{eq:def1_3} 
\end{align}
\end{definition}
\subsubsection*{Security under uniform message distribution}
The following definition extends Definition~\ref{def:dishonest_scheme} with a security requirement assuming that messages are independent and uniformly distributed.
\begin{definition}
\label{def:security_dishonest_uniform}
An $(n,\epsilon,N_1,N_2)$ scheme is said to be {\em secure against a
dishonest user under uniform message distribution}, if it guarantees decodability and security for an honest user even if the other user is dishonest (as defined above). That is, if $W_1$ and $W_2$ are independent and both are uniformly distributed, then when Bob 
 is honest, 
 \begin{align}
\max_{\s_2} \Pr\{\phi_1(Y_1^nS^{*n}) \neq W_1\} &< \epsilon\\
\max_{\s_2} I(W_1;Y_2^nS^{n}\Theta_2) &< \epsilon \label{eq:secrecy_dishonest_uniform_1}
\end{align}
are satisfied, and when Calvin is honest, 
\begin{align}
\max_{\s_1}\Pr\{\phi_2(Y_2^nS^{*n}) \neq W_2\} &< \epsilon\\
\max_{\s_1}I(W_2;Y_1^nS^{n}\Theta_1) &< \epsilon \label{eq:secrecy_dishonest_uniform_2}.
 \end{align}
 are satisfied. The maxima are taken over all adversarial acknowledging strategies.
\end{definition}

\begin{definition}
\label{def:rate_region_dishonest_uniform}
%TODO find a notation
The rate region $\mathcal{R}^2_{uDH}\subset \R_+^2$ is the set of rate pairs for which for every \mbox{$\epsilon >0$} there exists an $(n,\epsilon,N_1,N_2)$ scheme that is secure against a dishonest user under uniform message distribution and satisfies
\begin{align}
R_1-\epsilon<\frac{1}{n}N_1\  \text{and}\  R_2-\epsilon<\frac{1}{n}N_2.
\end{align}
\end{definition}
\subsubsection*{Distribution independent security} Below, we relax the assumption on the message distribution. This leads to a stronger notion of security that we call distribution independent security.
\begin{definition}
\label{def:security_dis}
An $(n,\epsilon,N_1,N_2)$ scheme is said to provide \emph{distribution independent security}, if it guarantees decodability and security for the honestly acknowledging user (or users) independently of the joint distribution $P_{W_1,W_2}$ of $(W_1,W_2)$.  That is, if Bob is honest,
 \begin{align}
 \max_{P_{W_1,W_2},\s_2} \Pr\{\phi_1(Y_1^nS^{*n}) \neq W_1\} &< \epsilon\\
\max_{P_{W_1,W_2},\s_2} I(W_1;Y_2^nS^{n}\Theta_2|W_2) &< \epsilon \label{eq:secrecy_dis_1}
\end{align}
are satisfied, and if Calvin is honest, then
\begin{align}
\max_{P_{W_1,W_2},\s_1}\Pr\{\phi_2(Y_2^nS^{*n}) \neq W_2\} &< \epsilon\\
\max_{P_{W_1,W_2},\s_1}I(W_2;Y_1^nS^{n}\Theta_1|W_1) &< \epsilon \label{eq:secrecy_dis_2}.
 \end{align}
 are satisfied. 
\end{definition}
\begin{definition}
\label{def:rate_region_dis}
The rate region $\mathcal{R}^2_{DIS}\subset \R_+^2$ is the set of rate pairs for which for every \mbox{$\epsilon >0$} there exists an $(n,\epsilon,N_1,N_2)$ scheme that provides distribution independent security and satisfies
\begin{align}
R_1-\epsilon<\frac{1}{n}N_1\  \text{and}\  R_2-\epsilon<\frac{1}{n}N_2.
\end{align}
\end{definition}
\subsubsection*{Security against a user with side information}
In the following two definitions we assume that a dishonest user can choose the  message distribution of the other user, but not his own.
\begin{definition}
\label{def:security_dishonest}
An $(n,\epsilon,N_1,N_2)$ scheme is said to be {\em secure against a
dishonest user with side information}, if it guarantees decodability and security for an honest user even if the other user is dishonest (as defined above) and can choose the message distribution of the honest user. That is, if $W_1$ and $W_2$ are independent, then if $W_2$ is uniformly distributed and  Bob is honest,
 \begin{align}
\max_{P_{W_1},\s_2} \Pr\{\phi_1(Y_1^nS^{*n}) \neq W_1\} &< \epsilon\\
\max_{P_{W_1},\s_2} I(W_1;Y_2^nS^{n}\Theta_2) &< \epsilon \label{eq:secrecy_dishonest_1}
\end{align}
are satisfied, whereas if $W_1$ is uniformly distributed and Calvin is honest, 
\begin{align}
\max_{P_{W_2},\s_1}\Pr\{\phi_2(Y_2^nS^{*n}) \neq W_2\} &< \epsilon\\
\max_{P_{W_2},\s_1}I(W_2;Y_1^nS^{n}\Theta_1) &< \epsilon \label{eq:secrecy_dishonest_2}
 \end{align}
 are satisfied. The maxima are taken over all adversarial acknowledging strategies and all possible message distributions of the honest user.
\end{definition}

\begin{definition}
\label{def:rate_region_dishonest}
The rate region $\mathcal{R}^2_{DH}\subset \R_+^2$ is the set of rate pairs for which for every \mbox{$\epsilon >0$} there exists an $(n,\epsilon,N_1,N_2)$ scheme that is secure against a dishonest user with side information and satisfies
\begin{align}
R_1-\epsilon<\frac{1}{n}N_1\  \text{and}\  R_2-\epsilon<\frac{1}{n}N_2.
\end{align}
\end{definition}

\subsection{Non-secure 1-to-$K$ broadcast}
Before summarizing our results, we restate the result from \cite{Marios2013,Wang12} that characterizes $\mathcal{R}^K$ in the known cases. Let $\pi$ denote a permutation of $\{1,2,\dots,K\}$ and $\pi_i$ the $i$th element of the permutation.
%TODO correct citation

\begin{theorem}
\label{thm:1-to-k}
For $K\leq3$ or for a symmetric channel with $K>3$ or for a one-sidedly fair rate tuple \mbox{$(R_1,\dots,R_K)\in \R_+^K$} with $K>3$, the capacity region $\mathcal{R}^K$ of the \mbox{1-to-$K$} broadcast erasure channel with state-feedback is characterized by the following inequality:
\begin{align}
\max_{\pi} \sum_{i=1}^K\frac{R_{\pi_i}}{1-\prod_{k=1}^i\delta_{\pi_k}}\leq 1  \label{eq:non_sec_region},
\end{align}
where the maximization is taken over all permutations $\pi$ of $\{1,\dots,K\}$.
\end{theorem}
Further, it was shown in  \cite{Wang12} and \cite{Marios2013} that \eqref{eq:non_sec_region} is an outer-bound for $\mathcal{R}^K$ in all cases.
%TODO correct citations everywhere
\begin{theorem}
\label{thm:non_sec_outerbound}
Any rate tuple \mbox{$(R_1,\dots,R_K)\in \R_+^K$} in $\mathcal{R}^K$ satisfies \eqref{eq:non_sec_region}.
\end{theorem}

\section{Summary of results}

In this section we provide an overview of the results that we present in this paper.

\subsection{Honest-but-curious users}
Our main result for honest-but-curious users is the characterization of the secret-message capacity region $\mathcal{R}^K_H$ for sending private messages to $K$ receivers over a broadcast erasure channel, for all the cases where the capacity region without secrecy constraints $\mathcal{R}^K$ has been characterized, namely, the $2$-user, $3$-user, symmetric $K$-user and one-sidedly fair $K$-user cases.

For all the mentioned cases, when the capacity region $\mathcal{R}^K$ is known, we prove the following theorem which describes the corresponding secret-message capacity region $\mathcal{R}^K_H$.

\begin{theorem}
\label{thm:1-to-k_sec}
For $K\leq3$ or for a symmetric channel with $K>3$ or for a one-sidedly fair rate tuple \mbox{$(R_1,\dots,R_K)\in \R_+^K$} with $K>3$, the secret-message capacity region $\mathcal{R}^K_H$ as defined in Definition~\ref{def:rate_region_honest} is characterized by the following inequality:
\ifonecolumn
\begin{align}
\max_{j\in\{1,\dots,K\}}\frac{R_j(1-\frac{\prod_{k=1}^K\delta_k}{\delta_j})}{(1-\delta_j)\frac{\prod_{k=1}^K\delta_k}{\delta_j}(1-\prod_{k=1}^K\delta_k)}+
\max_{\pi} \sum_{i=1}^K\frac{R_{\pi_i}}{1-\prod_{k=1}^i\delta_{\pi_k}}\leq 1 ,\label{eq:region_honest}
\end{align}
\else
\begin{multline}
\max_{j\in\{1,\dots,K\}}\frac{R_j(1-\frac{\prod_{k=1}^K\delta_k}{\delta_j})}{(1-\delta_j)\frac{\prod_{k=1}^K\delta_k}{\delta_j}(1-\prod_{k=1}^K\delta_k)}+\\
\max_{\pi} \sum_{i=1}^K\frac{R_{\pi_i}}{1-\prod_{k=1}^i\delta_{\pi_k}}\leq 1 ,\label{eq:region_honest}
\end{multline}
\fi
where the second maximization is taken over all permutations $\pi$ of $\{1,\dots,K\}$.
\end{theorem}
We prove the achievability part of Theorem~\ref{thm:1-to-k_sec} constructively by describing a linear scheme that achieves any rate tuple in $\mathcal{R}^K_H$ in the mentioned cases. The scheme together with the proof of its properties are given in Section~\ref{sec:honest_scheme}.

We also develop a converse proof to show that the scheme is optimal. Our converse proof inherently provides a new proof of Theorem~\ref{thm:non_sec_outerbound}. We provide the converse proof in Section~\ref{sec:converse}, which completes the proof of Theorem~\ref{thm:1-to-k_sec}.

\begin{figure}
\begin{center}
{\includegraphics[width=\figurewidth]{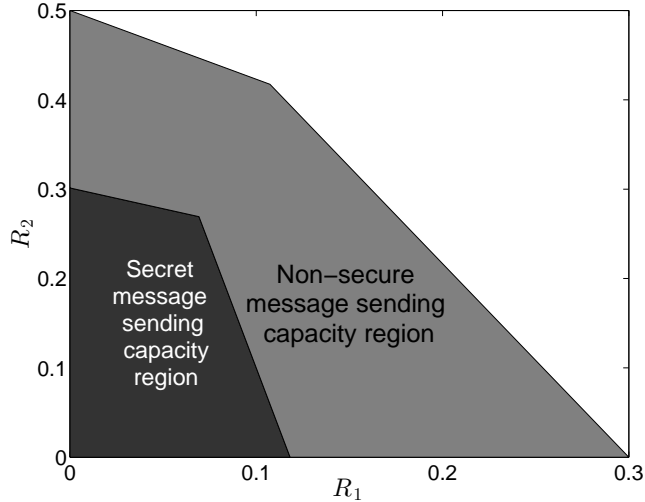}}
\end{center}
\caption{Non-secure message sending and secret-message sending capacity regions for $K=2$, $\delta_1=0.7$, $\delta_2=0.5$.}
\label{fig:sec_region}
\end{figure}

Comparing regions $\mathcal{R}^K$ and $\mathcal{R}^K_H$, the first term in \eqref{eq:region_honest} can be interpreted as the overhead for security. Indeed, in the scheme that we present, there is a key generation phase whose duration is proportional to this term. In \figurename~\ref{fig:sec_region}, we visualize this overhead for some specific parameter values.

\subsection{Dishonest users}
For the case of a dishonest user, we characterize the rate regions $\mathcal{R}^2_{uDH}, \mathcal{R}^2_{DH}$. We focus on security against a dishonest user with side information as defined in Definitions~\ref{def:security_dishonest}-\ref{def:rate_region_dishonest}. In particular, we show that the same rates are achievable against a dishonest user with side information as against honest-but-curious users, {\em i.e.,} $\mathcal{R}^2_{H}=\mathcal{R}^2_{DH}$. This implies $\mathcal{R}^2_{uDH}=\mathcal{R}^2_{DH}$, hence our result on $\mathcal{R}^2_{DH}$ implicitly characterizes $\mathcal{R}^2_{uDH}$ as well. We provide a formal description and proof for $K=2$, but the same ideas extend for $K>2$. The following theorem states that $\mathcal{R}^2_{DH}=\mathcal{R}^2_H$.
\begin{theorem}
\label{thm:dishonest}
The rate region $\mathcal{R}^2_{DH}$ as defined in Definition~\ref{def:rate_region_dishonest} is the set of all rate pairs $(R_1,R_2)\in \mathbb{R}_+^2$ which satisfy the following two inequalities:
\begin{align}
\frac{R_1(1-\delta_2)}{\delta_2(1-\delta_1)(1-\delta_1\delta_2)}+\frac{R_1}{1-\delta_1}+\frac{R_2}{1-\delta_1\delta_2}&\leq
1, \label{eq:th_dishonest_1}
\\ \frac{R_2(1-\delta_1)}{\delta_1(1-\delta_2)(1-\delta_1\delta_2)}+\frac{R_1}{1-\delta_1\delta_2}+\frac{R_2}{1-\delta_2}&\leq
1. \label{eq:th_dishonest_2}
\end{align}
\end{theorem}
%\begin{corollary}
%The region defined in Theorem~\ref{thm:dishonest} also characterizes the rate pairs of all possible schemes which provide semantic security.
%\end{corollary}
It is clear that $\mathcal{R}^2_{DH}\subseteq \mathcal{R}^2_H$, since the converse developed for the honest-but-curious case provides a valid outer bound. To prove that the region given by \eqref{eq:th_dishonest_1}-\eqref{eq:th_dishonest_2} is achievable, we construct a linear scheme that is secure against dishonest users and achieves any pair in the region. The scheme is described in Section~\ref{sec:dishonest_scheme}.

Theorem~\ref{thm:dishonest} gives a complete characterization of the problem considering security against a dishonest user with side information. Regarding distribution independent security we do not have such a characterization. We construct a scheme that satisfies this stronger security definition, however its optimality is not clear. The next theorem gives the rate region achieved by our scheme. 

\begin{theorem}
\label{thm:dis}
If a rate pair $(R_1,R_2)$ satisfies
\ifonecolumn
\begin{align}
\frac{R_1(1-\delta_2)}{\delta_2(1-\delta_1)(1-\delta_1\delta_2)}+\frac{R_2(1-\delta_1)}{\delta_1(1-\delta_2)(1-\delta_1\delta_2)}+\frac{R_1}{1-\delta_1}+\frac{R_2}{1-\delta_1\delta_2}&\leq
1, \label{eq:th2_1}
\\ \frac{R_1(1-\delta_2)}{\delta_2(1-\delta_1)(1-\delta_1\delta_2)}+\frac{R_2(1-\delta_1)}{\delta_1(1-\delta_2)(1-\delta_1\delta_2)}+\frac{R_1}{1-\delta_1\delta_2}+\frac{R_2}{1-\delta_2}&\leq
1. \label{eq:th2_2}
\end{align}
\else
\begin{multline}
\frac{R_1(1-\delta_2)}{\delta_2(1-\delta_1)(1-\delta_1\delta_2)}+\frac{R_2(1-\delta_1)}{\delta_1(1-\delta_2)(1-\delta_1\delta_2)}\\+\frac{R_1}{1-\delta_1}+\frac{R_2}{1-\delta_1\delta_2}\leq
1, \label{eq:th2_1}
\end{multline}
\vspace{-4mm}
\begin{multline}
\frac{R_1(1-\delta_2)}{\delta_2(1-\delta_1)(1-\delta_1\delta_2)}+\frac{R_2(1-\delta_1)}{\delta_1(1-\delta_2)(1-\delta_1\delta_2)}\\+\frac{R_1}{1-\delta_1\delta_2}+\frac{R_2}{1-\delta_2}\leq
1. \label{eq:th2_2}
\end{multline}
\fi
then $(R_1,R_2)\in \mathcal{R}^2_{DIS}$.
\end{theorem}
%\begin{corollary}
%The region of rate pairs in Theorem~\ref{thm:dis} support distribution independent semantic security.
%\end{corollary}

From the definitions it is clear that $\mathcal{R}^2_{DIS} \subseteq \mathcal{R}^2_{DH}$. We conjecture that there is a fundamental gap between $\mathcal{R}^2_{DIS}$ and $\mathcal{R}^2_{DH}$,  and $\mathcal{R}^2_{DIS} \subset \mathcal{R}^2_{DH}$ holds, but we leave the proof an open question. The scheme that constructively proves Theorem~\ref{thm:dis} is given in Section~\ref{sec:dis}.

\subsubsection*{Security against an eavesdropper}
Consider the special case when $K=2$ and $R_2=0$. There is only one receiver with nonzero rate and we aim to secure his message against the other, dishonest party. In this setting the other receiver is equivalent to a passive eavesdropper who overhears the communication. Note that the sender does not trust the feedback from the second receiver, so this feedback is simply ignored. In other words, in this particular setting there is no difference between giving potentially dishonest feedback and not giving any feedback at all. In the end, we have a broadcast channel with one receiver and an eavesdropper against whom we aim to secure a message. 

In the light of the argument above, the following definition naturally defines secret-message capacity against an eavesdropper.
\begin{definition}
The secret-message capacity $\mathcal{C}_{E}$ of a broadcast erasure channel with state-feedback against an eavesdropper is the largest $R$ for which $(R,0)\in \mathcal{R}^2_{DH}$.
\end{definition}
The following corollary characterizes $\mathcal{C}_{E}$. The result directly follows from Theorem~\ref{thm:dishonest}.

\begin{corollary}
\label{cor:single_capacity}
The secret-message capacity of a broadcast erasure channel with state-feedback against an eavesdropper is
\begin{align}
 \mathcal{C}_E=(1-\delta)\delta_E\frac{1-\delta\delta_E}{1-\delta\delta_E^2},\label{eq:single_capacity}
\end{align}
where $\delta_E$ denotes the erasure probability of the eavesdropper and $\delta$ that of the legitimate receiver.
\end{corollary}

From Wyner's result \cite{Wyner75} it is well known that the secret-message capacity of the same setting but without any feedback from the receiver is $(\delta_E-\delta)^+$. Our result shows how feedback helps to increase the achievable rate of secure communication.

Corollary~\ref{cor:single_capacity} also reveals the subtle difference between the secret-message sending and the secret-key generation problem. It was shown in \cite{Maurer1993,JDFPA10} that in the same setting, a key that is secret from the eavesdropper can be established between the sender and the legitimate receiver at rate $\delta_E(1-\delta)$, but the key is not known in advance by any of the parties. If we ask that the sender specifies in advance the secret which becomes a shared secret between the sender and the receiver after the protocol run, we arrive to the secret-message sending problem. From \eqref{eq:single_capacity} it is clear that the rate $\delta_E(1-\delta)$ is not achievable in this case. 

\begin{figure}
\begin{center}
{\includegraphics[width=\figurewidth]{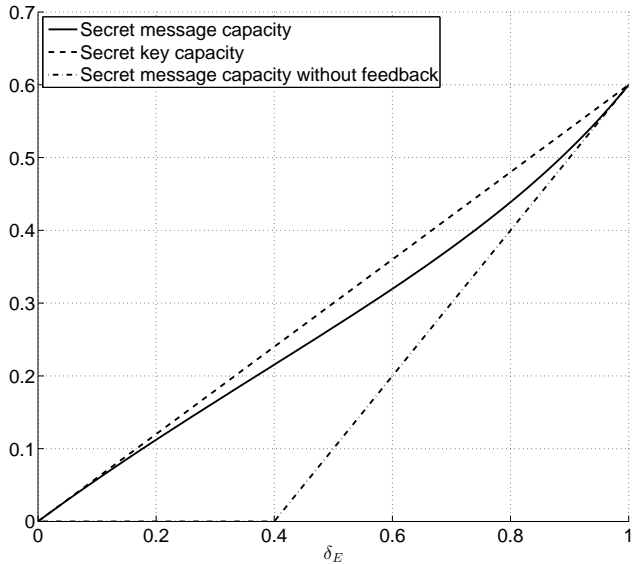}}
\end{center}
\caption{Secret-message and secret-key capacities with and without state-feedback for $\delta=0.4$.}
\label{fig:itw}
\end{figure}

As a comparison, on \figurename~\ref{fig:itw} we plot the secret-message capacity of the broadcast erasure channel against an eavesdropper with and without feedback from the receiver as well as the secret-key capacity of the same setting. Note that without feedback there is no difference between the secret-message sending and the secret-key generation problem.

\section{Honest-but-curious users}
\label{sec:honest_scheme} 
We prove the direct part of Theorem~\ref{thm:1-to-k_sec} by constructing a secure scheme against honest-but-curious users. At a high level, our scheme consists of two phases:
\begin{enumerate}
\item \emph{Key generation.} We create $K$ pairwise keys, each key is shared between Alice and one of the receivers, and it is perfectly secure from all the other receivers even if they collude.
\item \emph{Encrypted broadcast.} Using the keys set up in the first phase, we employ an encrypted version of the non-secure 1-to-$K$ broadcast scheme as we describe shortly.
\end{enumerate}

In our second phase, we build on a modified version of the linear scheme presented in \cite{Wang12} that achieves $\mathcal{R}^K$ as stated in Theorem~\ref{thm:1-to-k}. We refer to this scheme as the \emph{non-secure} 1-to-$K$ achievability scheme. Conceptually,  this algorithm has two main steps:
\begin{enumerate}[Step (a):]
\item Alice repeats each message packet $W_{1,1},\dots W_{1,N_1}\dots W_{K,N_K}$ until at least one of the receivers correctly receives it. We call $j$ the \emph{intended} receiver of a message packet $W_{j,i}$.
\item Alice sends linear combinations of the packets that are not received by their intended receiver in Step~(a). 
\end{enumerate}
A key contribution of \cite{Wang12} is in specifying how to construct the linear combinations in Step~(b) -- we refer the reader to \cite{Wang12} for the exact constructions, and highlight here the two important properties that we rely on: 
\begin{itemize}
\item A message packet  successfully delivered to its intended receiver in Step~(a) is never used in Step~(b).
\item The scheme achieves the rate points within $\mathcal{R}^K$ as stated in Theorem~\ref{thm:1-to-k}.
\end{itemize}

\subsection{Example}

\begin{table*}[t]
%\vspace{-.2cm}
\caption{An example of the protocol run.}
%\vspace{-.4cm}
\label{tab:example}
\center
      \begin{tabular}{p{2.3cm}lcccccc}
	\toprule
	&\multirow{2}{*}{Alice sends}&			Bob's&		Calvin's&		\multirow{2}{*}{Bob's key}&	\multirow{2}{*}{Calvin's key}& \multirow{2}{*}{Bob decoded} &\multirow{2}{*}{Calvin decoded}\\
	&		&		ACK&		ACK\\
	\midrule
 	\multirow{3}{*}{\parbox{2cm}{Key generation} $\begin{cases}\\\\\end{cases}$}  & $X_1$ random&	$\checkmark$&	$\times$&      $K_{B,1}=X_1$&		\\
			& $X_2$ random&	$\checkmark$&	$\checkmark$&      $K_{B,1}$&		\\
			& $X_3$ random&	$\times$&	$\checkmark$&      $K_{B,1}$&	$K_{C,1}=X_3$&	\\
			\midrule
 %\dotfill&\dotfill&\dotfill&\dotfill&\dotfill&\dotfill &\dotfill&\dotfill   \\
	\multirow{5}{*}{\parbox{2cm}{Encrypted message transmission} $\begin{cases}\\\\\\\\\end{cases}$}&  $X_4=W_{1,1}\oplus K_{B,1}$& $\times$ &$\checkmark$ & &$K_{C,1}$& \\			      
			      &  $X_5=W_{2,1}\oplus K_{C,1}$& $\times$ &$\checkmark$ & &$K_{C,1}$& &$W_{2,1}$\\ 
			      &  $X_{6}=W_{2,2}\oplus K_{C,1}$& $\times$ &$\times$ & &$K_{C,1}$& &$W_{2,1}$\\ 
			      &  $X_{7}=X_{6}$& $\checkmark$ &$\times$ & && &$W_{2,1}$\\
			      &  $X_{8}=X_{4}\oplus X_{7}$& $\checkmark$ &$\checkmark$ & & &  $W_{1,1}$&$W_{2,1},W_{2,2}$\\ 
\bottomrule
\end{tabular}      
\vspace{-.4cm}
\end{table*}

Before giving the detailed description of our scheme we show a small example which is suitable to highlight the ideas we use to build our protocol. 

Consider a setting with $K=2$. For convenience, we call the sender Alice, and the two receivers Bob and Calvin. In our example, Alice wants to securely send $N_1=1$ message packet  $W_1=[W_{1,1}]$ to Bob and $N_2=2$ message packets $W_{2}=[W_{2,1},W_{2,2}]$ to Calvin. The example protocol run is found in Table~\ref{tab:example}.

\noindent\emph{Key generation:}
\begin{enumerate}[(a)]

\item Alice transmits random (independent and uniformly distributed) packets $X_1,X_2,X_3$.  At the end of this phase, Alice and Bob share a secret key packet $K_{B,1}=X_1$ that Bob received and Calvin did not. Similarly, Alice and Calvin share the secret key packet $K_{C,1}=X_3$. The packet $X_2$ which was received by both Bob and Calvin is discarded.
\end{enumerate}

\noindent\emph{Encrypted message transmissions:}
\begin{enumerate}[(a)]
\setcounter{enumi}{1}

\item Alice secures Bob's first message packet with a one-time pad (using the secret key generated above) and repeatedly transmits an encrypted packet until either Bob or Calvin receives. In our example, this happens immediately ($X_4$). The packet received only by Calvin is a side information which enables us to efficiently use the channel at a later point.
\item In the next few transmissions ($X_5$-$X_{7}$) we do the same with Calvin's packets. As we see, if only Calvin receives ($X_5$), a part of the message is successfully delivered, however the key used for encryption can be used again securely to encrypt the next message packet ($X_6$). If neither Bob nor Calvin receives ($X_6$), the packet is simply repeated ($X_7$).
\item Once Bob also has a side information ($X_7$) packet, we send the sum of the two side information packets thereby sending information that is useful simultaneously for both receivers. This happens at transmission $X_{8}=X_{4}\oplus X_{7}$, where both Bob and Calvin can decode a novel message packet ($X_4$ is for Bob, $X_7$ is for Calvin). Note that at this step we do not need any new keys to secure the transmission.
\end{enumerate}
Through this small example we see the following important features of the scheme:
\begin{itemize}
\item The number of key packets we set up and consume is  smaller than the number of message packets we convey per user, because we can reuse certain keys if no other receiver has seen any packet encrypted with the given key.
\item We exploit side information packets that users have about each other's message to make a single transmission useful for both, without consuming  any new key.
\end{itemize}

\subsection{Detailed description}
We need to define a few parameters. The length of the secret keys we aim to set up for receiver $j$ (expressed in terms of packets) is $k_j$, and the length of the key generation phase in terms of transmissions is $n_1$. We define
\ifonecolumn
\begin{align}
k_j &= N_j\frac{1-\frac{\prod_{k=1}^K\delta_k}{\delta_j}}{1-\prod_{k=1}^K\delta_k} + \left(N_j\frac{1-\frac{\prod_{k=1}^K\delta_k}{\delta_j}}{1-\prod_{k=1}^K\delta_k}\right)^{3/4}, \mbox{ and\quad}
n_1 = \max_{j}  \frac{k_j+k_j^{3/4}}{(1-\delta_j)\frac{\prod_{k=1}^K\delta_k}{\delta_j}}.\label{eq:honest_parameters}
\end{align}
\else
\begin{align}
k_j &= N_j\frac{1-\frac{\prod_{k=1}^K\delta_k}{\delta_j}}{1-\prod_{k=1}^K\delta_k} + \left(N_j\frac{1-\frac{\prod_{k=1}^K\delta_k}{\delta_j}}{1-\prod_{k=1}^K\delta_k}\right)^{3/4}, \mbox{ and }\\
% k_1 &= N_1\frac{1-\delta_2\delta_3}{1-\delta_1\delta_2\delta_3} + \left(N_1\frac{1-\delta_2\delta_3}{1-\delta_1\delta_2\delta_3}\right)^{3/4} \\
% k_2 &= N_2\frac{1-\delta_1\delta_3}{1-\delta_1\delta_2\delta_3} + \left(N_2\frac{1-\delta_1\delta_3}{1-\delta_1\delta_2\delta_3}\right)^{3/4} \\
% k_3 &= N_3\frac{1-\delta_1\delta_2}{1-\delta_1\delta_2\delta_3} + \left(N_3\frac{1-\delta_1\delta_2}{1-\delta_1\delta_2\delta_3}\right)^{3/4} \\
n_1 &= \max_{j}  \frac{k_j+k_j^{3/4}}{(1-\delta_j)\frac{\prod_{k=1}^K\delta_k}{\delta_j}}.\label{eq:honest_parameters}
\end{align}
\fi
\subsection*{1) Key generation}
Let $K_j$ denote the key between Alice and receiver $j$.

Alice transmits $n_1$ random packets $X_1,\dots,X_{n_1}$ generated uniformly at random over \FqL. $K_j$ is the vector of the first $k_j$ packets $X_i$ for which $S_i=j$. If there are less than $k_j$ such packets, we stop and declare an error for receiver~$j$. 
In other words, Alice transmits random packets, and we treat a packet received by only one receiver as a shared secret between Alice and that receiver.

\subsection*{2) Encrypted broadcast}
\label{eq:}
We now follow the two transmission steps in the non-secure protocol, with the following modifications: in Step~(a), we encrypt the message packets using key packets as we specify in the following; in Step (b), we simply reuse the already encrypted packets from Step~(a) to create the required linear combinations -- we do not use additional key packets.

%\begin{enumerate}
Step (2.a):\;\; Before transmitting  each message packet to receiver $i$, Alice encrypts it by XOR-ing it with a key packet that has either not been used for encryption in the past, or if used, none of the other users received the corresponding packet.

Consider the transmissions to receiver $j$.
Initially, Alice encrypts the first packet for $j$ as  $W_{j,1}\oplus K_{j,1}$ and transmits it until it is received by at least one of the receivers. If only receiver $j$ receives this encrypted packet, she reuses the same key packet $K_{j,1}$ to encrypt the next message packet.  Subsequently, if for some $i$ and $\ell<N_1$, $k<k_1$: $X_i=W_{j,\ell}'=W_{j,\ell}\oplus K_{j,k}$, then
\begin{align}
X_{i+1}=\begin{cases} X_{i}, & \text{if $S_{i}=\emptyset$} \\ 
		      W_{1,\ell+1}'=W_{j,\ell+1}\oplus K_{j,k}, &\text{if $S_i=j$} \\
		      W_{1,\ell+1}'=W_{j,\ell+1}\oplus K_{j,k+1}, &\text{otherwise}.
\end{cases}
\end{align}
In other words, a key is reused until a packet encrypted using it is received by any other receiver. We declare an error if the $k_j$ key packets are not sufficient to encrypt all the $N_j$ message packets of $W_j$. Alice proceeds similarly for the other keys and messages.

Step (2.b):\;\; At the end of Step~(2.a), the receivers have received encrypted packets that are not intended for them as side information. We use the same encoding as in  Step~(b) of the non-secure protocol to  deliver these packets to their intended receivers.
%\end{enumerate}

\subsection{Analysis of  the secure protocol}
\label{sec:analysis_honest}
We need to show that conditions \eqref{eq:ch_inputs}-\eqref{eq:secrecy} are all satisfied. Condition (\ref{eq:ch_inputs}) is obviously satisfied by construction. We show the other required properties for receiver $j$, the same arguments apply to any $j$.

\subsubsection*{Security} We  first argue that our scheme satisfies (\ref{eq:secrecy}).
From construction,  at the end of the first phase we create a key $K_j$ with
\begin{align}
I(K_j;Y_1^{n_1},\dots,Y_{j-1}^{n_1},Y_{j+1}^{n_1},\dots,Y_K^{n_1}S^{n_1}) &=0. \label{eq:key_sec}
\end{align}
In Step (2.a), every packet $W_{j,\ell}'$ that any of the other receivers {\em receive} has been encrypted using a different key packet $K_{j,i}$; these key packets, from (\ref{eq:key_sec}), are secret from Calvin and David. Thus the packets received by the $K-1$ other receivers together are one-time pad encrypted and hence perfectly secret to them, even if they collude.
In Step~(2.b),  Alice transmits linear combinations of packets $W_{j,\ell}'$ that have not been received by receiver $j$, but have already been received by at least one of the other $K-1$ receivers. Thus, assuming these receivers collude, they do not receive any innovative $W_{j,\ell}'$. This concludes our argument and shows
\begin{align}
I(W_j;Y_1^n,\dots,Y_{j-1}^n,Y_{j+1}^n,\dots,Y_K^nS^n)=0.
\end{align}

\subsubsection*{Decodability} We next prove  (\ref{eq:decodability}).
Trivially, if no error is declared, receiver $j$ can retrieve $W_j$ from $W_j'$ using his key $K_j$. We show that the probability of declaring an error can be made arbitrarily small. It is enough to consider the following two error events since the other error events are similar: 
\begin{inparaenum}[(\it i)]
\item we do not obtain $k_j$ key packets for receiver~$j$ during the first phase, and \item $k_j$ key packets are not sufficient in Step~(2.a).
\end{inparaenum}

\noindent(\emph{i}) Let  $\kappa$ denote the number of packets in the first phase that are received only by receiver~$j$. Then, $\kappa$ is the sum of $n_1$ i.i.d.~Bernoulli variables drawn from $\Ber(p)$, where $p=(1-\delta_j)\frac{\prod_{k=1}^K\delta_k}{\delta_j}$. Thus,
$$\Expect{\kappa}=n_1p=n_1(1-\delta_j)\frac{\prod_{k=1}^K\delta_k}{\delta_j}\geq k_j+k_j^{3/4}.$$
The probability of error event (\emph{i}) equals 
\ifonecolumn
\begin{align}
\Prob{\kappa<k_j}&\leq\Prob{\Expect{\kappa}-\kappa>k_j^{3/4}}\leq\Prob{|\Expect{\kappa}-\kappa|>k_j^{3/4}}\leq e^{-c\sqrt{k_j}},
\end{align}
\else
\begin{align}
\Prob{\kappa<k_j}&\leq\Prob{\Expect{\kappa}-\kappa>k_j^{3/4}}\\
&\leq\Prob{|\Expect{\kappa}-\kappa|>k_j^{3/4}}\leq e^{-c\sqrt{k_j}},
\end{align}
\fi
for some constant $c>0$. The last inequality follows from the Chernoff-Hoeffding bound \cite{Hoeffding1963}. Selecting $N_1$ sufficiently large, this error probability can be made arbitrarily small.

\noindent(\emph{ii}) This error event is similar, it occurs if the number of packets that only Bob receives is significantly less than its expected value, and the same technique can be applied  to bound the probability of error. %to show that this error probability also can be made arbitrarily small. 
With this, we have shown that the scheme is secure against honest-but-curious users as defined in Definition~\ref{def:secrecy}. 

\subsubsection*{Rate of the scheme}  Finally, as for \eqref{eq:rates}, a straightforward calculation with the given parameters together with the capacity achieving property of non-secure 1-to-$K$ protocol  shows that our proposed scheme achieves any rate tuple within the region given by (\ref{eq:region_honest}). For completeness, we provide the rate calculation in Appendix~\ref{app:rate_calcualtion}. This concludes the proof of the achievability part of Theorem~\ref{thm:1-to-k_sec}.

\section{Dishonest users}
\label{sec:dishonest_scheme}

We consider the case when $K=2$ and one of the receivers potentially acknowledges dishonestly. The security of the scheme that we presented in the previous section crucially relies on honest feedback from \emph{all} receivers. If we want to provide security against dishonest users, then the secrecy of message $W_j$ should rely only on the acknowledgment of receiver $j$. The scheme we describe in this section provides this property. 

\subsection{Principles}
The structure of the new scheme follows the two-phase structure described previously. However, when we create a key for user $j$ or when we send an encrypted packet to him, instead of the feedback of the other user, we rely on the expected behavior channel. Interestingly, this does not require a sacrifice in rate as long as messages are independent and the message distribution of the dishonest user is uniform.

For illustration, consider the key generation phase. Assume that Alice transmits three random packets $X_1,X_2,X_3$, and assume Bob receives $X_1,X_2$, while Calvin receives $X_2,X_3$ as seen in our example in Table~\ref{tab:example}. If we cannot rely on Bob's and Calvin's honesty, but we do know that Bob and Calvin have received at most 2 packets each, we could allocate  $K_{1}=X_1\oplus X_2$ as the key between Alice and Bob, and $K_{2}=X_2\oplus X_3$ as the key shared by Alice and Calvin. Note that the number of such linear combinations that we can securely produce is the same as number of key packets that we could set up assuming honest feedback.

We can exploit the channel behavior also in the second phase such that we still have the property that the number of key packets needed is less then the number of message packets to secure. Assume now that Bob has a key $K_1=[K_{1,1},K_{1,2}]$. When we send encrypted packets to Bob, assume we expect Calvin to receive two out of three such transmissions -- but we do not know which two.  We then create three linear combinations of Bob's keys, say $K'_{1,1}=K_{B,1}$, $K'_{1,2}=K_{B,2}$, $K'_{1,3}=K_{B,1}\oplus K_{B,2}$, and transmit
$W_{1,1}\oplus K_{1,1}'$, $W_{1,2}\oplus K_{1,2}'$, and $W_{1,3}\oplus K_{1,3}'$. No matter which two of these Calvin receives the message remains secret.  Our protocol builds on these ideas.

\subsection{Detailed description}
Here we give the details of both phases. We observe that the dishonest user can deny the reception of side information packets by which he can hinder the use of XOR-ed transmissions. The honest user must not experience any decrease in rate even in that case. We limit the length of each step of the scheme to ensure this property.

The operation of the protocol utilizes a set of parameters which can be directly calculated before the protocol starts, and whose use will be described in the following.   
\begin{flalign}
k_B&=N_1\frac{1-\delta_2}{1-\delta_1\delta_2} +
\raisedtothreefourths{\left(N_1\frac{1-\delta_2}{1-\delta_1\delta_2}\right)}
\qquad
\\k_C&=N_2\frac{1-\delta_1}{1-\delta_1\delta_2}+\raisedtothreefourths{\left(N_2\frac{1-\delta_1}{1-\delta_1\delta_2}\right)}\label{eq:params-first}\\
 k_1&=\frac{k_B}{\delta_2} + \frac{1}{\delta_2}
\left(\frac{2k_B}{\delta_2}\right)^{3/4}\quad
 k_2=\frac{k_C}{\delta_1} + \frac{1}{\delta_1}
\left(\frac{2k_C}{\delta_1}\right)^{3/4}\\
n_1&=\max\Bigg\{\frac{k_1}{1-\delta_1}+\raisedtothreefourths{\left(\frac{k_1}{1-\delta_1}\right)},\ 
              \frac{k_2}{1-\delta_2}+\raisedtothreefourths{\left(\frac{k_2}{1-\delta_2}\right)}\Bigg\}\label{eq:key_gen_length}\\
% \end{align}
% \begin{align}
n_{2,1} &= \frac{N_1}{1-\delta_1\delta_2}+ \left(\frac{N_1}{1-\delta_1\delta_2}\right)^{3/4} \\
n_{2,2} &= \frac{N_2}{1-\delta_1\delta_2}+ \left(\frac{N_2}{1-\delta_1\delta_2}\right)^{3/4} \\
n_{2,3}'&= \frac{N_1}{1-\delta_1}+\left(\frac{N_1}{1-\delta_1}\right)^{3/4} - n_{2,1} \\
n_{2,3}'' &= \frac{N_2}{1-\delta_2}+\left(\frac{N_2}{1-\delta_2}\right)^{3/4} - n_{2,2} \\
%n_2'' &= \frac{N_2}{1-\delta_2}+\frac{N_1}{1-\delta_1\delta_2} +\left(\frac{N_2}{1-\delta_2}+\frac{N_2}{1-\delta_1\delta_2}\right)^{3/4} \\
n&=n_1+n_{2,1}+n_{2,2}+\max\left\{n_{2,3}',n_{2,3}''\right\}.\label{eq:params-last}
\end{flalign}

Using our protocol, Alice attempts to send $N_1$ message packets $W_1=(W_{1,1},\dots,{W}_{1,N_1})$ to Bob and $N_2$ message packets 	$W_2=(W_{2,1},\dots,{W}_{2,N_2})$ to Calvin using at most $n$ packet transmissions.  We show in Section~\ref{sec:analysis} that the probability she fails to do so can be made arbitrarily small.  
She proceeds in two steps.\\
{\em Key Generation}
\begin{enumerate}
\item Alice transmits $n_1$ packets $X_1,\ldots,X_{n_1}$ generated uniformly at random. 
\item If Bob receives less than $k_1$ packets we declare a protocol error for him. Similarly, an error is declared for Calvin if he receives less than $k_2$ packets. When an error is declared for both users, the protocol terminates. If not, we continue with the user not in error, as if the user in error did not exist.
\item  Let $X_1^B$  be a $L\times k_1$  matrix that has as columns the first $k_1$ packets that Bob acknowledged.  Alice and Bob create $k_B$ secret key packets as
%\begin{align}
$K_B=X^B_{1}G_{K_B},$
%\end{align}
where $G_{K_B}$ is a $(k_1\times k_B)$ matrix and is a parity check matrix of a $(k_1,k_1-k_B)$ maximum distance separable (MDS) code \cite{Sloane}.
Similarly, using the first $k_2$ packets that  Calvin acknowledges, Alice and Calvin create $k_C$ secret key packets using the matrix $G_{K_C}$. Matrices $G_{K_B}, G_{K_C}$ are publicly known and fixed in advance.

\end{enumerate}
{\em Message encryption and transmission}\\
{\em Encryption}
\begin{enumerate}
\setcounter{enumi}{3}
\item  Alice and Bob produce $N_1$ linear combinations of their $k_B$ secret key packets
as $K'_B=K_BG_{K'_{B}}$, where $G_{K'_{B}}$ is a
$(k_B\times N_1)$ matrix and is a generator matrix of an $(N_1,k_B)$ MDS
code which is also publicly known. Similarly, Alice and Calvin create $N_2$ 
linear combinations of their $k_C$ key packets.
\item Alice creates $N_1$ encrypted messages to send to Bob  
\begin{align}
U_{B,i}=W_{1,i}\oplus K'_{B,i}, \quad i=1\ldots N_1,
\end{align}
where $\oplus$ is addition in the $\mathbb{F}_q^L$ vector space. Let $U_B$
denote the set of $U_{B,i},i=1,\ldots,N_1$. She similarly produces
a set $U_C$ of $N_2$ encrypted messages to send to Calvin
\begin{align}
U_{C,i}=W_{2,i}\oplus K'_{C,i}, \quad i=1\ldots N_2.
\end{align}
\end{enumerate}
{\em Encrypted transmissions}
\begin{enumerate}
\setcounter{enumi}{5}
\item  Alice sequentially takes the first encrypted packet from $U_{B,i}$, $i=1\ldots N_1$, that is not yet acknowledged by either Bob or Calvin and repeatedly transmits it, until it is acknowledged by either receiver. That is,  if 
at time $i$ Alice transmits $X_i = U_{B,j}$ for some $j<N_1$, then 
\begin{align}
 X_{i+1}=\begin{cases}
X_{i}, & \text{if $S^\ast_{i}=\emptyset$}
\\ U_{B,j+1}, &\text{otherwise}.
\end{cases}
\end{align}
Alice continues these transmissions until all packets from $U_B$ are acknowledged or $n_{2,1}$ transmissions are already made in this step. In the former case, she continues with the next step. In the latter case, if Bob does not acknowledge $\frac{N_1(1-\delta_1)}{1-\delta_1\delta_2}$ packets, then he is considered to be dishonest and Alice continues with sending only Calvin's packets  using ARQ. Similarly, if Calvin does not acknowledge $\frac{N_1(1-\delta_2)}{1-\delta_1\delta_2}$ packets, then he is considered to be dishonest and Alice continues with sending only Bob's packets. In case neither receiver is considered to be dishonest, still $U_B$ is not completely delivered, Alice stops and an error is declared for both receivers.

\item Similarly, Alice sends not-yet-acknowledged  encrypted packets from $U_{C,i}$, $i=1\dots N_2$, until either Bob or Calvin acknowledges. If at time $i$ Alice transmits $X_i=U_{C,j}$ for some $j<N_2$, then
\begin{align}
 X_{i+1}=\begin{cases}
X_{i}, & \text{if $S^\ast_{i}=\emptyset$}
\\ U_{C,j+1}, &\text{otherwise}.
\end{cases}
\end{align}
Alice continues these transmissions until all packets from $U_C$ are acknowledged or $n_{2,2}$ transmissions are already made in this step. In the former case, she continues with the next step. In the latter case, if Bob does not acknowledge $\frac{N_2(1-\delta_1)}{1-\delta_1\delta_2}$ packets, then he is considered to be dishonest and Alice continues with sending only Calvin's packets  using ARQ. Similarly, if Calvin does not acknowledge $\frac{N_2(1-\delta_2)}{1-\delta_1\delta_2}$ packets, then he is considered to be dishonest and Alice continues with sending only Bob's packets. In case neither receiver is considered to be dishonest, still $U_C$ is not completely delivered, Alice stops and an error is declared for both receivers.
\item 
Let $Q_B$ denote the set of packets that only Calvin acknowledged in Step~6. Similarly, $Q_C$ denotes those packets that only Bob acknowledged in Step~7. Alice sequentially takes packets from $Q_B$ and $Q_C$. For each transmission, she takes the first packet from $Q_B$ that Bob has not acknowledged together with the first packet from $Q_C$ that Calvin has not yet acknowledged and she transmits the XOR of the two packets. 
If at time $i$ Alice transmits $X_i=Q_{B,j}\oplus Q_{C,\ell}$ for some $j<|Q_B|, \ell<|Q_C|$, then
\begin{align}
X_{i+1}=\begin{cases}
X_i, &\text{if $S^\ast_{i}=\emptyset$,} \\
Q_{B,j+1}\oplus Q_{C,\ell},&\text{if $S^\ast_{i}=\{1\}$,} \\
Q_{B,j}\oplus Q_{C,\ell+1},&\text{if $S^\ast_{i}=\{2\}$,} \\
Q_{B,j+1}\oplus Q_{C,\ell+1},&\text{if $S^\ast_{i}=\{1,2\}$.} 
\end{cases}
\end{align}
\end{enumerate}
Alice continues with the XOR-ed transmissions until either receiver acknowledges all his packets. If Bob has already acknowledged all packets from $Q_B$, Alice repeats packets that are not yet acknowledged by Calvin from $Q_C$. Similarly, if Calvin has already  acknowledged all packets from $Q_C$, then  Alice continues with repeating the remaining packets for Bob from $Q_B$.

If at any point, the overall number of transmissions would exceed $n$ as defined in \eqref{eq:params-last} we stop and declare an error for the party (or parties) who has not acknowledged all his encrypted message packets.

\subsection{Analysis}
\label{sec:analysis}
Below, we prove that the above scheme is secure against a dishonest user with side information
and runs without error with high probability. The rate assertion of the
theorem follows from a simple numerical evaluation with the given parameter
values. 

\subsubsection{Security}

In our argument we focus on the secrecy of $W_1$ against a dishonest
Calvin, but the same reasoning works for $W_2$ against a dishonest Bob as
well.  Since we do not intend to give security guarantees to a dishonest
user and consider at most one user to be dishonest, we may assume that Bob
is honest. Moreover, under our definition of dishonest user with side information, $W_1$ and
$W_2$ are independent and the latter is uniformly distributed over its
alphabet, but the distribution of $W_1$ is arbitrary and controlled by the
dishonest Calvin.

To analyze the secrecy of $W_1$, we may, without loss of generality, assume
that no error was declared for Bob during the key generation phase. Recall
that an error is declared for Bob only if Bob fails to acknowledge at least
$k_1$ packets. If an error was in fact declared for Bob, no information
about Bob's message $W_1$ is ever transmitted by Alice.
%\footnote{More precisely, if $E_\text{I-B}$ is the indicator random variable for an error being declared for Bob in the key generation phase, \[ I(W_1;Y_2^n,S^n,\Theta_2) \leq I(W_1;Y_2^n,S^n,\Theta_2,E_\text{I-B}) = I(W_1;Y_2^n,S^n,\Theta_2|E_\text{I-B}) \leq I(W_1;Y_2^n,S^n,\Theta_2|E_\text{I-B} = 0).\] To avoid clutter, we leave out the conditioning event in the rest of this subsection.}. 
However, note that we do account for this error event when we analyze the probability of error for Bob in the Section~\ref{sec:prob-of-error}.

We need to show that the secrecy condition \eqref{eq:secrecy_dishonest_1} is satisfied by the scheme, {\em i.e.,}~Bob's message remains secret from Calvin even if Calvin controls the distribution of $W_1$, and applies any acknowledging strategy. In the proof we omit taking the maximum, but the argument holds for any $P_{W_1}$ and for any adversarial strategy, so the statement follows. 

We use the following three lemmas. 
\begin{lemma}
\label{lem:secret key generation}
When Bob is honest and no error is declared for Bob in the
key generation phase,
\begin{align}
I(K_B;Y_2^{n_1}S^{n_1})\leq k_Be^{-c_1\sqrt{k_1}},\label{eq:secret
key leakage}
\end{align}
if $k_1= \frac{k_B}{\delta_2} +
\frac{1}{\delta_2}\raisedtothreefourths{\left(\frac{2k_B}{\delta_2}\right)}
$ and $k_B\geq \frac{2}{\delta_2}$,
where $c_1>0$ is some constant. Moreover, $K_B$ is uniformly distributed over its alphabet.
\end{lemma}
Lemma~\ref{lem:secret key generation} shows that $I(K_B;Y_2^{n_1}S^{n_1})$ can be made small, {\em
  i.e.,}~the key generation phase is secure. The key facts we use in proving this lemma are \begin{inparaenum}[(\it i)] 
                                               \item the number of packets seen by Calvin concentrates around its mean and 
                                               \item an MDS parity check matrix can be used to perform privacy amplification in the packet erasure setting.
                                               \end{inparaenum} The proof is delegated to Appendix~\ref{sec:secret_key_lemma}.

Let ${1}_{B,i}^C$ be the indicator random variable for the
event that Calvin observes the packet $U_{B,i}$ either in its pure
form or in a form where the $U_{B,i}$ packet is added with some
$U_{C,j}$ packet. Let $M_B^C$ be the random variable which denotes the
number of distinct packets of $U_B$ that Calvin observes, so $M_B^C =
\sum_{i=1}^{N_1} 1_{B,i}^C.$ Given this notation, we have the following lemmas:
\begin{lemma}
\label{lem:analysis:first-entropy-term}
$H(Y_2^n|Y_2^{n_1}S^n\Theta_2U_C) 
\leq \Expect{M_B^C} .$
\end{lemma}
We prove Lemma~\ref{lem:analysis:first-entropy-term} in Appendix~\ref{sec:first_entropy_term}.
\ifonecolumn
\begin{lemma}
\label{lem:analysis:second-entropy-term}
$H(Y_2^n|W_1Y_2^{n_1}S^n\Theta_2U_C)\geq
\Expect{\min\left(k_B,M_B^C\right)} -
I(K_B;Y_2^{n_1}S^{n_1}).$
\end{lemma}
\else
\begin{lemma}
\label{lem:analysis:second-entropy-term}
\begin{multline}
H(Y_2^n|W_1Y_2^{n_1}S^n\Theta_2U_C)\\\geq \Expect{\min\left(k_B,M_B^C\right)} -
I(K_B;Y_2^{n_1}S^{n_1}).
\end{multline}
\end{lemma}
\fi
We prove Lemma~\ref{lem:analysis:second-entropy-term} in Appendix~\ref{sec:second_entropy_term}.

Using the results of Lemmas~\ref{lem:secret key generation}-\ref{lem:analysis:second-entropy-term}, we conclude the proof as follows. We have that 
\ifonecolumn
\begin{align}
I(W_1;Y_2^nS^n\Theta_2) \leq I(W_1;Y_2^nS^n\Theta_2U_C)
 = I(W_1;Y_2^n|Y_2^{n_1}S^n\Theta_2U_C), \label{eq:security-analysis1}
\end{align}
\else
\begin{align}
I(W_1;Y_2^nS^n\Theta_2) &\leq I(W_1;Y_2^nS^n\Theta_2U_C)\\
 &= I(W_1;Y_2^n|Y_2^{n_1}S^n\Theta_2U_C), \label{eq:security-analysis1}
\end{align}
\fi
where the last equality used the fact that $\Theta_A,\Theta_2,W_2,S^n$ are independent of $W_1$ and we may express $Y_2^{n_1},U_C$ as deterministic
functions of $\Theta_A,\Theta_2,W_2,S^n$. We use Lemmas~\ref{lem:analysis:first-entropy-term}-\ref{lem:analysis:second-entropy-term} in \eqref{eq:security-analysis1}, to get
\ifonecolumn
\begin{align}
I(W_1;Y_2^nS^n\Theta_2) \leq \Expect{\max\left(0,M_B^C-k_B\right)} 
+ I(K_B;Y_2^{n_1}S^{n_1}). \label{eq:security-analysis2}
\end{align}
\else
\begin{multline}
I(W_1;Y_2^nS^n\Theta_2) \leq \\\Expect{\max\left(0,M_B^C-k_B\right)} 
+ I(K_B;Y_2^{n_1}S^{n_1}). \label{eq:security-analysis2}
\end{multline}
\fi
Lemma~\ref{lem:secret key generation} gives a bound for the second
term. We bound the first term using concentration inequalities, in particular, we use the Chernoff-Hoeffding bound \cite{Hoeffding1963} for the purpose. In
order to do this, let $Z_{B,i}$ be the number of repetitions of a
packet $U_{B,i}$ that Alice makes until Bob acknowledges it (where we
count both the transmission in pure form and in addition with some
packet from $U_C$). Note that the random variables $Z_{B,i}$ are
independent of each other and have the same distribution.  This
follows from the fact that the $S_i$ sequence is i.i.d., and each
$S_i$ is independent of $(Y_2^{i-1},S^{i-1},\Theta_2)$. In other
words, Calvin can exert no control over the channel state. Further,
for the same reason, with every repetition the chance that Calvin
obtains the transmission is $1-\delta_2$. This implies that the
indicator random variables $1_{B,i}^C$ are i.i.d. with
\begin{align} \Prob{1_{B,i}^C=1} = (1-\delta_2) + \delta_1\delta_2(1-\delta_2) +
\ldots = \frac{1-\delta_2}{1-\delta_1\delta_2}.\end{align}
Notice that $M_B^C$ is a sum of $N_1$ such independent random variables, and
hence
$\Expect{M_B^C} = N_1\frac{1-\delta_2}{1-\delta_1\delta_2}.$
Since
$k_B = N_1\frac{1-\delta_2}{1-\delta_1\delta_2} +
\raisedtothreefourths{\left(N_1\frac{1-\delta_2}{1-\delta_1\delta_2}\right)},
$
by applying Chernoff-Hoeffding bound we have
\ifonecolumn
\begin{align}
\Expect{\max\left(0,M_B^C-k_B\right)} &\leq N_1 \Prob{M_B^C>k_B} \leq N_1
e^{-c_2\sqrt{N_1}}, \label{eq:ch_bound}
\end{align}
\else
\begin{align}
\Expect{\max\left(0,M_B^C-k_B\right)} &\leq N_1 \Prob{M_B^C>k_B} \\&\leq N_1
e^{-c_2\sqrt{N_1}}, \label{eq:ch_bound}
\end{align}
\fi
for a constant $c_2>0$. Substituting this back to \eqref{eq:security-analysis2} and using Lemma~\ref{lem:secret key generation}, we get
\begin{align}
I(W_1;Y_2^nS^n\Theta_2) \leq N_1e^{-c_2\sqrt{N_1}} + k_Be^{-c_2\sqrt{k_B}},
\end{align}
for constants $c_1,c_2>0$. By choosing a large enough value
of $N_1$, we may meet \eqref{eq:secrecy_dishonest_1}\footnote{Recall from
\eqref{eq:params-first}-\eqref{eq:params-last} that by saying that we choose
$N_1$ large enough we cause $n$ to be large enough.}.

\subsubsection{Error probability}\label{sec:prob-of-error}
We need to bound the probability that an error is declared for
Bob. If there is no error for Bob, he will be able to decode $W_1$. An error happens if \begin{inparaenum}[\it (i)]
                                                                                        \item Bob receives less than $k_1$ packets in the first phase, or 
                                                                                        \item he does not receive enough encrypted message packets in Steps~6 and~8 before the protocol terminates.
                                                                                        \end{inparaenum}
These error events have the same nature as the error events of our scheme for honest-but-curious users. An error happens if Bob collects significantly fewer packets than he is expected to receive in a particular step. The probability of these events can be made arbitrarily small by applying the same technique as in Section~\ref{sec:analysis_honest}. We omit details to avoid repetitive arguments.

A straightforward computation using the parameters in \eqref{eq:params-first}-\eqref{eq:params-last} shows that the achieved rate region matches the region claimed in Theorem~\ref{thm:dishonest}. We give the calculation in Appendix~\ref{app:rate_calcualtion}.

\subsubsection{Complexity considerations}
It is clear from the analysis in Section~\ref{sec:analysis} that the
length $n$ of the scheme grows as
$\max\{O(\log^2(\frac{1}{\epsilon}),O(\frac{1}{{\epsilon'}^4}))\}$, where
$\epsilon$ is the security and probability of error parameter, and 
$\epsilon'$ is the gap parameter associated with the rate. The algorithmic complexity is quadratic
in $n$; quadratic from the matrix multiplication to produce the key.

\subsection{Distribution independent scheme}
\label{sec:dis}
In the following we describe a scheme which satisfies distribution independent security as defined in Definition~\ref{def:security_dis}. Before that we provide the intuition behind the construction. The protocol in Section~\ref{sec:dishonest_scheme} cannot satisfy distribution independent security, because of the following. 
Assume Calvin knows his message a-priori and he acknowledges dishonestly in the key generation phase. Then, $K_C$ is constructed of packets that Calvin does not know, but in the second phase, Alice uses this $K_C$ to encrypt his message. If Calvin acknowledges honestly in the second phase, then he learns $K_C$ from $U_C$, because he already knows $W_2$. Since $K_C$ is a linear combination of key generation packets, this way Calvin might learn (in expectation) a $n_1(1-\delta_2)+k_C$ dimension subspace from the space spanned by the key generation packets. The expected number of key generation packets that either Bob or Calvin receives is $n_1(1-\delta_1\delta_2)$ and for a large $n$, $n_1(1-\delta_2)+k_C+k_B> n_1(1-\delta_1\delta_2)$, hence $K_B$ is not independent of Calvin's observation, which means that Bob's key is not secure.

We can overcome this issue if we modify the key generation phase and make sure that no packet used in generating Calvin's key is contributing Bob's key, thus  $U_C$ is conditionally independent of Bob's key given Calvin's observation of the protocol and $W_2$. This results in two separate key generation phases, one for Bob and one for Calvin.

\subsubsection{Protocol description}
We need two parameters for determining the number of key generation packets.
\begin{align}
n_{1,1}=\frac{k_1}{1-\delta_1}+\raisedtothreefourths{\left(\frac{k_1}{1-\delta_1}\right)}, \quad
n_{1,2}=\frac{k_2}{1-\delta_2}+\raisedtothreefourths{\left(\frac{k_2}{1-\delta_2}\right)}.
\end{align}
We use all other parameters and notations as introduced in Section~\ref{sec:dishonest_scheme}.

\noindent{\em Key Generation}
\begin{enumerate}
\item Alice transmits $n_{1,1}$ uniformly random packets $X_1,\ldots,X_{n_1}$ independent of $W_1$, $W_2$. From the first $k_1$ packets  that Bob acknowledges (matrix $X^B_1$) Alice computes Bob's key as $K_B=X^B_{1}G_{K_B}$. If Bob does not acknowledge $k_1$ packets we declare an error for him.
\item Alice transmits $n_{1,2}$ uniformly random packets $X_{n_1+1},\ldots,X_{n_1+n_2}$ independent of $W_1$, $W_2$. From the first $k_2$ packets that Calvin acknowledges out of these $n_{1,2}$ packets (matrix $X^C_1$),  Alice computes Calvins's key as $K_C=X^C_{1}G_{K_C}$. If Calvin does not acknowledge $k_1$ packets we declare an error for him.
\item When an error is declared for both users, the protocol terminates. If not, we continue with the user not in error, as if the user in error did not exist.
\end{enumerate}
{\em Message encryption and transmission}\\
Steps 4-8 are exactly the same as described in Section~\ref{sec:dishonest_scheme}.

This scheme provides distribution independent security, which property is proved in Appendix~\ref{app:dis_sec_proof}. The proof follows the same lines as the analysis of Section~\ref{sec:analysis}. This, together with a straightforward rate calculation completes the proof of Theorem~\ref{thm:dis}.

\section{Converse}
\label{sec:converse}
We show the converse part of Theorem~\ref{thm:1-to-k_sec}, by which we conclude the proof of Theorems~\ref{thm:1-to-k_sec}, \ref{thm:dishonest} and Corollary~\ref{cor:single_capacity}. This result proves the optimality of the schemes presented in Section~\ref{sec:honest_scheme} and in Section~\ref{sec:dishonest_scheme}.
\begin{proof}
We present our proof for $K=3$, the generalization of the same argument for any $K$ is straightforward. We are going to show that for any~$j$ and any~$\pi$ 
\ifonecolumn
\begin{align}
\frac{R_j(1-\frac{\delta_1\delta_2\delta_3}{\delta_j})}{(1-\delta_j)\frac{\delta_1\delta_2\delta_3}{\delta_j}(1-\delta_1\delta_2\delta_3)}+\frac{R_{\pi_1}}{1-\delta_{\pi_1}}+\frac{R_{\pi_2}}{1-\delta_{\pi_1}\delta_{\pi_2}}+\frac{R_{\pi_3}}{1-\delta_{\pi_1}\delta_{\pi_2}\delta_{\pi_3}}\leq 1 \label{eq:no_max}
\end{align}
\else
\begin{multline}
\frac{R_j(1-\frac{\delta_1\delta_2\delta_3}{\delta_j})}{(1-\delta_j)\frac{\delta_1\delta_2\delta_3}{\delta_j}(1-\delta_1\delta_2\delta_3)}+\\
\frac{R_{\pi_1}}{1-\delta_{\pi_1}}+\frac{R_{\pi_2}}{1-\delta_{\pi_1}\delta_{\pi_2}}+\frac{R_{\pi_3}}{1-\delta_{\pi_1}\delta_{\pi_2}\delta_{\pi_3}}\leq 1 \label{eq:no_max}
\end{multline}
\fi
holds, which implies the statement of the theorem. Also, to avoid cumbersome notation we show (\ref{eq:no_max}) for $j=1$ and $\pi=(1,2,3)$. With simple relabeling, the same argument holds for any~$j$ and~$\pi$.
\ifonecolumn 
\begin{align} 
n&\geq \sumin H(X_i)\geq \sumin H(X_i|Y_1^{i-1}S^{i-1})=\sumin H(X_i|Y_1^{i-1}Y_2^{i-1}S^{i-1})+I(X_i;Y_2^{i-1}|Y_1^{i-1}S^{i-1})  \\
&= \sumin H(X_i|Y_1^{i-1}Y_2^{i-1}Y_3^{i-1}S^{i-1})+I(X_i;Y_2^{i-1}|Y_1^{i-1}S^{i-1})+I(X_i;Y_3^{i-1}|Y_1^{i-1}Y_2^{i-1}S^{i-1})\\
&= \sumin H(X_i|W_1W_2W_3Y_1^{i-1}Y_2^{i-1}Y_3^{i-1}S^{i-1}) \label{eq:term_1}\\
&\quad+I(X_i;Y_2^{i-1}|Y_1^{i-1}S^{i-1}) \label{eq:term_2}\\
&\quad+I(X_i;Y_3^{i-1}|Y_1^{i-1}Y_2^{i-1}S^{i-1}) \label{eq:term_3}\\
&\quad+I(X_i;W_1W_2W_3|Y_1^{i-1}Y_2^{i-1}Y_3^{i-1}S^{i-1})\label{eq:term_4}
\end{align}
\else
\begin{align} 
n&\geq \sumin H(X_i)\geq \sumin
H(X_i|Y_1^{i-1}S^{i-1})\\
&=\sumin H(X_i|Y_1^{i-1}Y_2^{i-1}S^{i-1})+I(X_i;Y_2^{i-1}|Y_1^{i-1}S^{i-1})  \\
&= \sumin H(X_i|Y_1^{i-1}Y_2^{i-1}Y_3^{i-1}S^{i-1})\\
&\quad+I(X_i;Y_2^{i-1}|Y_1^{i-1}S^{i-1}) \\
&\quad+I(X_i;Y_3^{i-1}|Y_1^{i-1}Y_2^{i-1}S^{i-1})\\
&= \sumin H(X_i|W_1W_2W_3Y_1^{i-1}Y_2^{i-1}Y_3^{i-1}S^{i-1}) \label{eq:term_1}\\
&\quad+I(X_i;Y_2^{i-1}|Y_1^{i-1}S^{i-1}) \label{eq:term_2}\\
&\quad+I(X_i;Y_3^{i-1}|Y_1^{i-1}Y_2^{i-1}S^{i-1}) \label{eq:term_3}\\
&\quad+I(X_i;W_1W_2W_3|Y_1^{i-1}Y_2^{i-1}Y_3^{i-1}S^{i-1})\label{eq:term_4}
\end{align}
\fi
In the following Lemmas~\ref{lemma:term_1}-\ref{lemma:term_4} we give bounds on each of the terms (\ref{eq:term_1})-(\ref{eq:term_4}).
%In the following  lemmas we give bounds on each of these terms. Lemmas~\ref{lemma:term_1}-\ref{lemma:term_4} give bounds on terms (\ref{eq:term_1})-(\ref{eq:term_4}). 
Combining these results together gives (\ref{eq:no_max}) and in turn the statement of the theorem. 
The detailed proofs of these lemmas are delegated to Appendix~\ref{app:converse_lemmas}.
\end{proof}
\subsection{Proof of Theorem~\ref{thm:non_sec_outerbound}}
\label{sec:non_sec_outerbound_proof}
\begin{proof}
It is sufficient to prove the inequality for $\pi=(1,2,3)$. By relabeling, the same argument holds for any $\pi$. By repeating the first steps of the previous proof and bounding term (\ref{eq:term_1}) by 0, we have
\begin{align} 
n&\geq \sumin I(X_i;Y_2^{i-1}|Y_1^{i-1}S^{i-1}) \label{eq:term_2b}\\
&\quad+I(X_i;Y_3^{i-1}|Y_1^{i-1}Y_2^{i-1}S^{i-1}) \label{eq:term_3b}\\
&\quad+I(X_i;W_1W_2W_3|Y_1^{i-1}Y_2^{i-1}Y_3^{i-1}S^{i-1})\label{eq:term_4b}
\end{align}
Lemmas~\ref{lemma:term_2}-\ref{lemma:term_4} give bounds on terms (\ref{eq:term_2b})-(\ref{eq:term_4b}) respectively. Combining these gives the stated inequality. 
\end{proof}

\subsection{Interpretation of the converse proof}
To facilitate understanding, beside our formal proof through Lemmas~\ref{lemma:term_1}-\ref{lemma:term_1_helper} here we provide some intuitive interpretation of terms \eqref{eq:term_1}-\eqref{eq:term_4} and of the inequalities we derive. It will be helpful to match terms to the steps of our scheme, but we stress that the proof holds for any possible scheme.

In Lemma~\ref{lemma:term_1} we see the following (here we omit small terms for simplicity):
\ifonecolumn
\begin{align}
(1-\delta_1)\delta_2\delta_3\sumin H(X_i|Y_1^{i-1}Y_2^{i-1}Y_3^{i-1}W_1W_2W_3S^{i-1}) \geq \frac{nR_1(1-\delta_2\delta_3)}{1-\delta_1\delta_2\delta_3}.
\end{align}
\else
\begin{align}
(1-\delta_1)\delta_2\delta_3\sumin H(X_i|Y_1^{i-1}Y_2^{i-1}Y_3^{i-1}W_1W_2W_3S^{i-1}) \\\geq \frac{nR_1(1-\delta_2\delta_3)}{1-\delta_1\delta_2\delta_3}.
\end{align}
\fi
The entropy term on the LHS of this inequality accounts for fresh randomness sent by the source. In or scheme we call this the key generation phase. The constant factor $(1-\delta_1)\delta_2\delta_3$ suggests that a random packet becomes a key for receiver~1 if only he receives the transmission. The RHS of the inequality corresponds to the expected number of (encrypted) $W_1$ packets that not only receiver~1 gets, but some other receivers also overhear. These are the packets that need to be secured, thus for perfect secrecy, receiver~1 needs at least the same number of key packets. This lower bound on term \eqref{eq:term_1} suggests that any scheme has to introduce some source randomness. We find it natural to call it key generation.

Terms~\eqref{eq:term_2}-\eqref{eq:term_4} correspond to the second phase of our protocol. Term \eqref{eq:term_4} corresponds to the first step of the message transmission phase (see Step (a)), when the sender ensures that the receivers {\em together} could decode all the messages. Terms~\eqref{eq:term_2}-\eqref{eq:term_3} account for the encoded transmissions. E.g.~\eqref{eq:term_2} intuitively corresponds to ``a packet that is of interest for receiver~1 and known by receiver~2''. Indeed, Lemma~\ref{lemma:term_2} lower bounds this term with the expected number of transmissions that are needed to convey  to receiver~1 the side information overheard by receiver~2.

\subsection{Lemmas}
Here we state the lemmas that we use in the converse proof. The proofs of the lemmas are found in Appendix~\ref{app:converse_lemmas}.
\begin{lemma}
\label{lemma:term_1}
From conditions (\ref{eq:ch_inputs})-(\ref{eq:rates_honest}) it follows that
\ifonecolumn
\begin{align}
\sumin H(X_i|Y_1^{i-1}Y_2^{i-1}Y_3^{i-1}W_1W_2W_3S^{i-1}) \geq \frac{nR_1(1-\delta_2\delta_3)}{(1-\delta_1)\delta_2\delta_3(1-\delta_1\delta_2\delta_3)}-\E_8, 
\end{align}
\else
\begin{multline}
\sumin H(X_i|Y_1^{i-1}Y_2^{i-1}Y_3^{i-1}W_1W_2W_3S^{i-1}) \\\geq \frac{nR_1(1-\delta_2\delta_3)}{(1-\delta_1)\delta_2\delta_3(1-\delta_1\delta_2\delta_3)}-\E_8,  
\end{multline}
\fi
where $\E_8=\E_7\frac{1-\delta_2\delta_3}{(1-\delta_1)\delta_2\delta_3}$, and
$\E_7$ is an error constant specified in Lemma~\ref{lemma:term_1_helper}.
\end{lemma}

\begin{lemma}
\label{lemma:term_2}
From conditions (\ref{eq:ch_inputs})-(\ref{eq:rates_honest}) it follows that
\begin{align}
\sumin I(X_i;Y_2^{i-1}|Y_1^{i-1}S^{i-1}) \geq \frac{nR_1}{1-\delta_1}-\frac{nR_1}{1-\delta_1\delta_2}-\E_1,
 \end{align}
where $\E_1= \frac{h_2(\epsilon)+\epsilon }{1-\delta_1}$.
\end{lemma}

\begin{lemma}
\label{lemma:term_3}
From conditions (\ref{eq:ch_inputs})-(\ref{eq:rates_honest}) it follows that
\ifonecolumn
\begin{align}
\sumin I(X_i;Y_3^{i-1}|Y_1^{i-1}Y_2^{i-1}S^{i-1}) \geq \frac{n(R_1+R_2)}{1-\delta_1\delta_2}-\frac{n(R_1+R_2)}{1-\delta_1\delta_2\delta_3}-\E_2,
 \end{align}
\else
\begin{multline}
\sumin I(X_i;Y_3^{i-1}|Y_1^{i-1}Y_2^{i-1}S^{i-1}) \geq \\\frac{n(R_1+R_2)}{1-\delta_1\delta_2}-\frac{n(R_1+R_2)}{1-\delta_1\delta_2\delta_3}-\E_2,
 \end{multline}
\fi
\end{lemma}
where $\E_2=\frac{h_2(2\epsilon)+2\epsilon }{1-\delta_1\delta_2}$.

\begin{lemma}
\label{lemma:term_4}
From conditions (\ref{eq:ch_inputs})-(\ref{eq:rates_honest}) it follows that
\ifonecolumn
\begin{align}
\frac{n(R_1+R_2+R_3)}{1-\delta_1\delta_2\delta_3}-\E_3\leq \sumin I(X_i;W_1W_2W_3|Y_1^{i-1}Y_2^{i-1}Y_3^{i-1}S^{i-1})\leq\frac{n(R_1+R_2+R_3)}{1-\delta_1\delta_2\delta_3}
 \end{align}
\else
\begin{multline}
\frac{n(R_1+R_2+R_3)}{1-\delta_1\delta_2\delta_3}-\E_3\\\leq \sumin I(X_i;W_1W_2W_3|Y_1^{i-1}Y_2^{i-1}Y_3^{i-1}S^{i-1}) \\\leq\frac{n(R_1+R_2+R_3)}{1-\delta_1\delta_2\delta_3}
 \end{multline}
\fi
where $\E_3= \frac{h_2(3\epsilon)+3\epsilon }{1-\delta_1\delta_2\delta_3}$.
\end{lemma}

\begin{lemma}
\label{lemma:balance}
From the definition of the channel it follows that
\ifonecolumn
\begin{align}
\sumin H(X_i|Y_1^{i-1}Y_2^{i-1}Y_3^{i-1}W_1W_2W_3S^{i-1})\geq \frac{1-\delta_2\delta_3}{(1-\delta_1)\delta_2\delta_3}\sumin I(X_i;Y_1^{i-1}|Y_2^{i-1}Y_3^{i-1}W_1W_2W_3S^{i-1})   
\end{align}
\else
\begin{multline}
\sumin H(X_i|Y_1^{i-1}Y_2^{i-1}Y_3^{i-1}W_1W_2W_3S^{i-1})\geq \\ \frac{1-\delta_2\delta_3}{(1-\delta_1)\delta_2\delta_3}\sumin I(X_i;Y_1^{i-1}|Y_2^{i-1}Y_3^{i-1}W_1W_2W_3S^{i-1})   
\end{multline}
\fi
\end{lemma}

\begin{lemma}
\label{lemma:security}
From the security condition (\ref{eq:secrecy}) it follows that
\begin{align}
\E_4 &> \sumin I(X_i;W_1|Y_2^{i-1}Y_3^{i-1}S^{i-1}),
 \end{align}
where $\E_4 = \frac{\epsilon}{1-\delta_2\delta_3}$.
\end{lemma}

\begin{lemma}
\label{lemma:term_1_helper}
From conditions (\ref{eq:ch_inputs})-(\ref{eq:secrecy}) it follows that
\ifonecolumn
\begin{align}
\sumin I(X_i;Y_1^{i-1}|Y_2^{i-1}Y_3^{i-1}S^{i-1}W_1W_2W_3) \geq \frac{nR_1}{1-\delta_1\delta_2\delta_3} -\E_7, 
\end{align}
\else
\begin{multline}
\sumin I(X_i;Y_1^{i-1}|Y_2^{i-1}Y_3^{i-1}S^{i-1}W_1W_2W_3) \geq \\\frac{nR_1}{1-\delta_1\delta_2\delta_3} -\E_7, 
\end{multline}
\fi
where  $\E_7=2\E_2'+\E_4+\E_5+\E_6$, $\E_{5}=\frac{h_2(\epsilon)+\epsilon }{1-\delta_2\delta_3}$, $\E_{6}=\frac{h_2(\epsilon)+\epsilon }{1-\delta_1\delta_2\delta_3}$, and  $\E_2'=\frac{h_2(2\epsilon)+2\epsilon }{1-\delta_2\delta_3}$.

\end{lemma}

\section{Discussion}
\newcommand{\Advssdis}{\ensuremath{{\mathbf{Adv}}^{\mathrm{ss}}_{\mathrm{dis}} }}
\newcommand{\Advss}{\ensuremath{{\mathbf{Adv}}^{\mathrm{ss}} }}
\newcommand{\Advds}{\ensuremath{{\mathbf{Adv}}^{\mathrm{ds}} }}
\newcommand{\Advssdiss}{\ensuremath{{\mathbf{Adv}}^{*\mathrm{ss}}_{\mathrm{dis}} }}
\newcommand{\Advdsdis}{\ensuremath{{\mathbf{Adv}}^{\mathrm{ds}}_{\mathrm{dis}} }}
\newcommand{\Advmisdis}{\ensuremath{{\mathbf{Adv}}^{\mathrm{mis}}_{\mathrm{dis}} }}
\newcommand{\Advmis}{\ensuremath{{\mathbf{Adv}}^{\mathrm{mis}} }}
\renewcommand{\S}{\mathcal{S}}
\label{sec:sec_defs}

\subsection{Channels with correlated erasures}
Our results can be generalized for memoryless channels with arbitrary correlation between the erasure events. We consider the case of honest-but-curious users. Let $\delta_{\mathcal{N}}$ denote the erasure probability that the set~$\mathcal{N}$ of receivers jointly experience and $p_j$ the probability that only user~$j$ receives:
\begin{align}
\delta_{\mathcal{N}}&=\Prob{\forall j\in \mathcal{N}: Y_{j,i}=\perp}\\
p_j &= \Prob{Y_{j,i}=X_{j,i}, \forall k\neq j: Y_{k,i}=\perp}.
\end{align}
Our outer bound proof relies on knowing the statistical behavior of the channel but not on its independence property. Using the parameters defined above, a straightforward generalization of the proof in Section~\ref{sec:converse} results in a more general bound: any rate tuple \mbox{$(R_1,\dots,R_K)\in \R_+^K$} in $\mathcal{R}^K_H$ satisfies 
\ifonecolumn
\begin{align}
\max_{j\in\{1,\dots,K\}}\frac{R_j(1-\delta_{\{-j\}})}{p_j(1-\delta_{\{1,\dots,K\}})}+
\max_{\pi} \sum_{i=1}^K\frac{R_{\pi_i}}{1-\delta_{\{\pi_1, \dots, \pi_i \}}}\leq 1 ,
\end{align}
\else
\begin{multline}
\max_{j\in\{1,\dots,K\}}\frac{R_j(1-\delta_{\{-j\}})}{p_j(1-\delta_{\{1,\dots,K\}})}+\\
\max_{\pi} \sum_{i=1}^K\frac{R_{\pi_i}}{1-\delta_{\{\pi_1, \dots, \pi_i \}}}\leq 1 ,
\end{multline}
\fi
where $\delta_{\{-j\}}$ is used as a shorthand for $\delta_{\{1,\dots,j-1,j+1,\dots,K\}}$. 

Our arguments for the key generation phase also do not exploit the independence property of the channel. It follows that for the broadcast erasure channel the upper bound on the achievable key generation rate for user~$j$
\begin{align}
\max I(X_i;Y_{j,i}|Y_{1,i},\dots,Y_{j-1,i},Y_{j+1,i},\dots,Y_{K,i})=p_j
\end{align}
derived in \cite{Maurer1993} is achievable without the requirement of independent erasures. In the parameter definitions and proofs  $p_j$ is substituted as the key generation rate for user~$j$. Consider our parameter definitions for honest-but-curios users in eq.~\eqref{eq:honest_parameters}. With this modification,
\ifonecolumn
\begin{align}
k_j &= N_j\frac{1-\delta_{\{-j\}}}{1-\delta_{\{1,\dots,K\}}} + \left(N_j\frac{1-\delta_{\{-j\}}}{1-\delta_{\{1,\dots,K\}}}\right)^{3/4}, \mbox{ and\quad}
n_1 = \max_{j}  \frac{k_j+k_j^{3/4}}{p_j}.
\end{align}
\else
\begin{align}
k_j &= N_j\frac{1-\delta_{\{-j\}}}{1-\delta_{\{1,\dots,K\}}} + \left(N_j\frac{1-\delta_{\{-j\}}}{1-\delta_{\{1,\dots,K\}}}\right)^{3/4}, \mbox{ and }\\
% k_1 &= N_1\frac{1-\delta_2\delta_3}{1-\delta_1\delta_2\delta_3} + \left(N_1\frac{1-\delta_2\delta_3}{1-\delta_1\delta_2\delta_3}\right)^{3/4} \\
% k_2 &= N_2\frac{1-\delta_1\delta_3}{1-\delta_1\delta_2\delta_3} + \left(N_2\frac{1-\delta_1\delta_3}{1-\delta_1\delta_2\delta_3}\right)^{3/4} \\
% k_3 &= N_3\frac{1-\delta_1\delta_2}{1-\delta_1\delta_2\delta_3} + \left(N_3\frac{1-\delta_1\delta_2}{1-\delta_1\delta_2\delta_3}\right)^{3/4} \\
n_1 &= \max_{j}  \frac{k_j+k_j^{3/4}}{p_j}.
\end{align}
\fi

The second phase of our achievability algorithm (Step~(2.b)) uses a capacity achieving non-secure coding scheme. For the cases where such a scheme is available (for non-secure coding schemes we refer the reader to \cite{Wang12}), our secure protocol naturally extends for a channel with correlated erasures.

\subsection{Security notions}

We formulate our results in information theoretic terms, defining
secrecy as a mutual information term being negligibly small. In the
realm of computational cryptography it is more common to prove
security of an encryption scheme by showing distinguishing security or
semantic security. To facilitate the interpretation of our results and
to allow a fair comparison with other schemes, we cite a recent result
from \cite{Vardy12}, which shows equivalence between the two
approaches. However, the definitions of distinguishing security and semantic security are applicable only for a single user setting. We can directly use these definitions for the special cases when $R_1=0$ or $R_2=0$, i.e., when we consider security against an eavesdropper. In these cases, our definition of security against a dishonest
user with side information is equivalent to semantic security. We extend the notion of semantic security such that it handles joint message distributions,
which results in a definition matching distribution independent security. We will give the definitions for Bob's security, the security for Calvin is completely symmetric.

It is common to define the \emph{advantage} of the adversary to express the gain that the adversary obtains by observing the protocol. 
Considering security against an eavesdropper, the adversarial advantage expressed in terms of mutual information (mis = mutual information security) is defined as:
\begin{align}
{\mathbf{Adv}}^{\text{mis}} = \max_{P_{W_1},\s_2} I(W_1;Y_2^nS^n\Theta_2).
\end{align}

The notion of semantic security captures the intuition that the probability that an adversary can compute a function $f$ of the message should not increase significantly after observing the protocol compared to the a-priori probability of a correct guess. The semantic security advantage is defined as
\ifonecolumn
\begin{align}
{\mathbf{Adv}}^{\text{ss}} = \max_{f,P_{W_1},\s_2} \left \{ \max_{\mathcal{A}} \Prob{\A(Y_2^n,S^n,\s_2)=f(W_1)}\right.\left. -\max_\S \Prob{\S(P_{W_1},f)=f(W_1)}  \right\},
\end{align}
\else
\begin{multline*}
{\mathbf{Adv}}^{\text{ss}} = \max_{f,P_{W_1},\s_2} \left \{ \max_{\mathcal{A}} \Prob{\A(Y_2^n,S^n,\s_2)=f(W_1)}\right.\\\left. -\max_\S \Prob{\S(P_{W_1},f)=f(W_1)}  \right\},
\end{multline*}
\fi
where $f$ is an arbitrary function of $W_1$, $\A$ is any function that the adversary
may compute after observing the protocol and $\S$ is a simulator
trying to compute $f$ without accessing the protocol output. Here
also, $W_2$ is uniformly distributed and independent of $W_1$. The
term simulator to denote guessing functions comes from the intuition
that ideally there exists an algorithm (simulator) that simulates the
run of a protocol without having access to the message and whose
output is indistinguishable from the output of a real
protocol.  Theorems~1,~5 and~8 from \cite{Vardy12} prove the following
inequalities:
\begin{align}
{\mathbf{Adv}}^{\text{ss}} \leq \sqrt{2\cdot{\mathbf{Adv}}^{\text{mis}}}; \ 
{\mathbf{Adv}}^{\text{mis}}\leq 4 \cdot{\mathbf{Adv}}^{\text{ss}} \log\left(\frac{2^n}{{\mathbf{Adv}}^{\text{ss}}}\right)
\end{align}
This result shows that requirement \eqref{eq:secrecy_dishonest_1} is naturally equivalent to semantic security. {\em i.e.,}~a small $\epsilon$ in \eqref{eq:secrecy_dishonest_1} implies that ${\mathbf{ Adv}}^{\text{ss}}$ is also small.

Applying the above definition for a case when $R_2>0$ implicitly assumes that Calvin cannot choose the distribution of his own message $W_2$. We now extend the definition of semantic security such that it does not rely on the distribution of $W_2$, which results in a stronger notion of security. We define the adversarial advantage for this case as
\ifonecolumn
\begin{align}
{\mathbf{Adv}}^{\text{ss}}_{\text{dis}}= \max_{f,P_{W_1,W_2},\s_2} \left \{ \max_{\mathcal{A}} \Prob{\A(Y_2^n,S^n,\s_2,W_2)=f(W_1,W_2)} \right. -\left.\max_\S \Prob{\S(P_{W_1,W_2},f,W_2)=f(W_1,W_2) } \right\} \label{eq:ss_dis}.
\end{align}
\else
\begin{multline}
{\mathbf{Adv}}^{\text{ss}}_{\text{dis}}=\\ \max_{f,P_{W_1,W_2},\s_2} \left \{ \max_{\mathcal{A}} \Prob{\A(Y_2^n,S^n,\s_2,W_2)=f(W_1,W_2)} \right. \\-\left.\max_\S \Prob{\S(P_{W_1,W_2},f,W_2)=f(W_1,W_2) } \right\} \label{eq:ss_dis}.
\end{multline}
\fi
Note that here we allow the simulator to have access to the message $W_2$ which an honest Calvin will learn. 
 The corresponding definition of adversarial advantage for distribution independent security directly comes from~\eqref{eq:secrecy_dis_1}:
 \begin{align}
 {\mathbf{Adv}}^{\text{mis}}_{\text{dis}} =\max_{P_{W_1,W_2},\s_2}I(W_1;Y_2^nS^n\Theta_2|W_2) 
 \end{align}
We show in Appendix~\ref{app:equivalence} the following lemma, which implies that requirement \eqref{eq:secrecy_dis_1} is equivalent to this extended notion of semantic security. 

\begin{lemma}
\label{lem:equivalence}
\begin{align}
{\mathbf{Adv}}^{\text{\em ss}}_{\text{\em dis}} &\leq \sqrt{2\cdot{\mathbf{Adv}}^{\text{\em mis}}_{\text{\em dis}}} \\
{\mathbf{Adv}}^{\text{\em mis}}_{\text{\em dis}}&\leq 4 \cdot{\mathbf{Adv}}^{\text{\em ss}}_{\text{\em dis}} \log\left(\frac{2^n}{{\mathbf{Adv}}^{\text{\em ss}}_{\text{\em dis}}}\right).
\end{align}
\end{lemma}

These results show that although security definitions might look quite different at first sight, there is no fundamental difference between these notions of security. As a corollary, our results also characterize the rate regions that are achievable by any scheme that provides semantic security.

\bibliographystyle{IEEEtran}
\bibliography{secrecy.bib}

\appendices

\section{Proof of lemmas in Section~\ref{sec:analysis}}
\subsection{Proof of Lemma~\ref{lem:secret key generation}}
\label{sec:secret_key_lemma}
\begin{proof}
With a slight abuse of notation, in the following $X_1^{BC}$ will denote
the \emph{actual} packets Calvin received (not necessarily the same as
those that he acknowledges) out of the first $k_1$ packets that Bob received.
Note that here we assume that an error was not declared for Bob in the key
generation phase and hence Bob did receive at least $k_1$ packets in the
key generation phase.  Also let $X_1^{B\emptyset}$ be the packets seen only
by Bob among the first $k_1$ he receives.  Let $I_{B\emptyset}$ and
$I_{BC}$ be the index sets corresponding to $X_1^{B\emptyset}$ and
$X_1^{BC}$. Recall that $X_1^B$ denotes the first $k_1$ packets received by
Bob. The notation $M^{I}$ will denote a matrix $M$ restricted to the
columns defined by index set $I$. Given this,
\ifonecolumn
\begin{align}
&I(K_B;Y_2^{n_1}S^{n_1})=I(X_1^BG_{K_B};X_1^{BC}S^n)\\&=H(X_1^BG_{K_B})-H(X_1^BG_{K_B}|X_1^{BC}S^n)
 \\ &=k_B-H(X_1^BG_{K_B}|X_1^{BC}S^n) \\
&=k_B-H(\left[X_1^{B\emptyset}G_{K_B}^{I_{B\emptyset}}\quad
X_1^{BC}G_{K_B}^{I_{BC}}\right]|X_1^{BC}S^n)\\
&=k_B - H(X_1^{B\emptyset}G_{K_B}^{I_{B\emptyset}}|X_1^{BC}S^n)\\
&=k_B-H(X_1^{B\emptyset}G_{K_B}^{I_{B\emptyset}}|S^n),
\end{align}
\else
\begin{align}
&I(K_B;Y_2^{n_1}S^{n_1})=I(X_1^BG_{K_B};X_1^{BC}S^n)\\&=H(X_1^BG_{K_B})-H(X_1^BG_{K_B}|X_1^{BC}S^n)
 \\ &=k_B-H(X_1^BG_{K_B}|X_1^{BC}S^n) \\
&=k_B-H(\left[X_1^{B\emptyset}G_{K_B}^{I_{B\emptyset}}\quad
X_1^{BC}G_{K_B}^{I_{BC}}\right]|X_1^{BC}S^n)\\
&=k_B - H(X_1^{B\emptyset}G_{K_B}^{I_{B\emptyset}}|X_1^{BC}S^n)\\
&=k_B-H(X_1^{B\emptyset}G_{K_B}^{I_{B\emptyset}}|S^n),
\end{align}
\fi
where the third equality follows from the MDS property of the matrix
$G_{K_B}$. Using the same property, we have
\begin{align}
&H(X_1^{B\emptyset}G_{K_B}^{I_{B\emptyset}}|S^n)=\sum_{i=0}^{k_1}\min\{i,k_B\}L\log
q\Prob{|X_1^{B\emptyset}|=i} \\ 
&\geq k_B\sum_{i=k_B}^{k_1}\Prob{|X_1^{B\emptyset}|=i} \\
&= k_B\Prob{|X_1^{B\emptyset}|\geq k_B} \\ 
&=k_B\left(1-\Prob{|X_1^{B\emptyset}|<k_B}\right) \\ 
&=k_B\left(1-\Prob{|X_1^{BC}|\geq k_1-k_B}\right) \\ 
&\stackrel{\text{(a)}}{\geq} k_B\left(1-\Prob{|X_1^{BC}|\geq
(1-\delta_2)k_1 + \raisedtothreefourths{k_1}}\right) \\ 
&\geq k_B
\left(1-\Prob{\left||X_1^{BC}|-E\left[|X_1^{BC}|\right]\right|>\raisedtothreefourths{k_1}}\right),
\end{align}
where the inequality (a) follows from the fact that the conditions
on $k_B$ and $k_1$ imply that 
\[k_1-k_B \geq (1-\delta_2)k_1 + \raisedtothreefourths{k_1}.\]
The Chernoff-Hoeffding bound gives that for some constant $c_1>0$
\begin{align}
\Prob{\left||X_1^{BC}|-E\left[|X_1^{BC}|\right]\right|>\raisedtothreefourths{k_1}}
\leq e^{-c_1\sqrt{k_1}}.
\end{align}
So, we have that 
\begin{align}
I(K_B;Y^{n_1}S^{n})&\leq k_Be^{-c_1\sqrt{k_1}}.
\end{align}
%which clearly can be made arbitrarily small with a sufficiently large
%$N_1$.

The final assertion of the lemma is a simple consequence of the MDS
property of the code and the fact that $X^{n_1}$ are uniformly i.i.d.
\end{proof}

\subsection{Proof of Lemma~\ref{lem:analysis:first-entropy-term}}
\label{sec:first_entropy_term}
\begin{proof}
Let $U_B^C$ be a vector of length $N_1$
such that the $i$-th element $U_{B,i}^C$ is $U_{B,i}$ if Calvin observes
this $U_{B,i}$ either in the pure form 
%during step 2
or added with some element of $U_C$, and $U^C_{B,i}=\perp$ otherwise.
%during step 4  of the encrypted message transmission phase.
Let $1_{B,i}^C$ is the indicator random variable for the event
$U_{B,i}^C\neq\perp$, so $M_B^C =
\sum_{i=1}^{N_1} 1_{B,i}^C.$
It is easy to see that the following are information equivalent
({\em i.e.,}~we can express each side as a deterministic function of the other)
\[ (Y_2^n,S^n,\Theta_2,U_C) \equiv (U_B^C,Y_2^{n_1},S^n,\Theta_2,U_C).\]
Therefore,
\[ H(Y_2^nS^n\Theta_2U_C) = H(U_B^CY_2^{n_1}S^n\Theta_2U_C).\]
\ifonecolumn
\begin{align}
&H(Y_2^n|Y_2^{n_1}S^n\Theta_2U_C) = H(U_B^C|Y_2^{n_1}S^n\Theta_2U_C)\\
&= \sum_{i=1}^{N_1} H(U_{B,i}^C|U_B^{C\,i-1}Y_2^{n_1}S^n\Theta_2U_C)\\
&= \sum_{i=1}^{N_1}
H(U_{B,i}^C|1_{B,i}^CU_B^{C\,i-1}Y_2^{n_1}S^n\Theta_2U_C)\\
&\leq \sum_{i=1}^{N_1} H(U_{B,i}^C|1_{B,i}^C)\\
&\leq \sum_{i=1}^{N_1} \Prob{1_{B,i}^C=1}= \Expect{\sum_{i=1}^{N_1} 1_{B,i}^C}.
\end{align}
\else
\begin{align}
&H(Y_2^n|Y_2^{n_1}S^n\Theta_2U_C) = H(U_B^C|Y_2^{n_1}S^n\Theta_2U_C)\\
&= \sum_{i=1}^{N_1} H(U_{B,i}^C|U_B^{C\,i-1}Y_2^{n_1}S^n\Theta_2U_C)\\
&= \sum_{i=1}^{N_1}
H(U_{B,i}^C|1_{B,i}^CU_B^{C\,i-1}Y_2^{n_1}S^n\Theta_2U_C)\\
&\leq \sum_{i=1}^{N_1} H(U_{B,i}^C|1_{B,i}^C)\\
&\leq \sum_{i=1}^{N_1}  \Prob{1_{B,i}^C=1}\\
&= \Expect{\sum_{i=1}^{N_1} 1_{B,i}^C} .
\end{align}
\fi
where the third equality follows from the fact that the indicator random variable
$1_{B,i}^C$ is a deterministic function of the conditioning
random variables.

%%%%%%%%%%%%% N.B. %%%%%%%%%%%%%%%%%%%%%%%%%%%
%\textcolor{red}{NB: In the proof above I have used notation conforming to
%the alternative protocol specification of Laszlo.}
\end{proof}

\subsection{Proof of Lemma~\ref{lem:analysis:second-entropy-term}}
\label{sec:second_entropy_term}
\begin{proof}
We adopt the notation for $U_B^C$ and $1_{B,i}^C$ introduced in the proof
of Lemma~\ref{lem:analysis:first-entropy-term}. In addition, let ${K'}_B^C$
be defined in a similar manner as $U_B^C$ such that ${K'}_{B,i}^C=\perp$ if
$U_{B,i}^C=\perp$ and ${K'}_{B,i}^C={K'}_{B,i}$ otherwise. Also, let
$1_B^{C}$ be the vector of indicator random variables $1_{B,i}^C$,
$j=1,\ldots,N_1$.

Proceeding as in the proof of 
Lemma~\ref{lem:analysis:first-entropy-term}, we have
\ifonecolumn
\begin{align}
&H(Y_2^n|W_1Y_2^{n_1}S^n\Theta_2U_C) 
= H(U_B^C|W_1Y_2^{n_1}S^n\Theta_2U_C)\\
&=
%H(U_B^C|W_1Y_2^{n_1}S^n\Theta_2U_C)\\ &= 
H({K'}_B^C|W_1Y_2^{n_1}S^n\Theta_2U_C)\geq H({K'}_B^C|1_B^CW_1Y_2^{n_1}S^n\Theta_2U_C)\\
&= H({K'}_B^C|1_B^C) - I({K'}_B^C;W_1Y_2^{n_1}S^n\Theta_2U_C|1_B^C)
\end{align}
\else
\begin{align}
&H(Y_2^n|W_1Y_2^{n_1}S^n\Theta_2U_C) 
= H(U_B^C|W_1Y_2^{n_1}S^n\Theta_2U_C)\\
&=
%H(U_B^C|W_1Y_2^{n_1}S^n\Theta_2U_C)\\ &= 
H({K'}_B^C|W_1Y_2^{n_1}S^n\Theta_2U_C)\\
&\geq H({K'}_B^C|1_B^CW_1Y_2^{n_1}S^n\Theta_2U_C)\\
&= H({K'}_B^C|1_B^C) - I({K'}_B^C;W_1Y_2^{n_1}S^n\Theta_2U_C|1_B^C)
\end{align}
\fi
But, from the MDS property of $G_{K'_B}$, and the fact that $K_B$ is
uniformly distributed over its alphabet, we have
\ifonecolumn
\begin{align}
H({K'}_B^C|1_B^C) &= \sum_{i=1}^{N_1}
\min(i,k_B)\Prob{\sum_{j=1}1_{B,j}^C=i} = \Expect{\min\left(k_B,\sum_{i=1}^{N_1} 1_{B,i}^C\right)}.
\end{align}
\else
\begin{align}
H({K'}_B^C|1_B^C) &= \sum_{i=1}^{N_1}
\min(i,k_B)\Prob{\sum_{j=1}1_{B,j}^C=i} \\
&= \Expect{\min\left(k_B,\sum_{i=1}^{N_1} 1_{B,i}^C\right)} .
\end{align}
\fi 
Also,
\ifonecolumn
\begin{align}
I({K'}_B^C;W_1Y_2^{n_1}S^n\Theta_2U_C|1_B^C)
&\stackrel{\text{(a)}}{=} 
I({K'}_B^C;Y_2^{n_1}S^{n_1}|1_B^C)
\leq I({K'}_B^C1_B^C;Y_2^{n_1}S^{n_1})
\leq I({K}_B;Y_2^{n_1}S^{n_1}).
\end{align}
\else
\begin{align}
I({K'}_B^C;W_1Y_2^{n_1}S^n\Theta_2U_C|1_B^C)
&\stackrel{\text{(a)}}{=} 
I({K'}_B^C;Y_2^{n_1}S^{n_1}|1_B^C)\\
&\leq I({K'}_B^C1_B^C;Y_2^{n_1}S^{n_1})\\
&\leq I({K}_B;Y_2^{n_1}S^{n_1}).
\end{align}
\fi 
where (a) follows from the fact that the distribution of $W_2$ (uniform and
independent of $S^n,\Theta_A,\Theta_2$) implies that $U_C$ is independent
of $\Theta_A,S^n$ and using this we can argue that 
the following is Markov chain
\[ {K'}_{B}^C  - (1_B^C,Y_2^{n_1},S^{n_1}) - (W_1,\Theta_2,U_C).\] 
By substituting back we obtain the claim of the lemma.
\end{proof}

\section{Rate calculation}
\label{app:rate_calcualtion}
\subsection{Honest-but-curious users}
The rate achieved for user $j$ is $R_j=\lim_{n\rightarrow\infty}\frac{N_j}{n}$. Compared to the non-secure 1-to-$K$ protocol we have an overhead of $n_1$ transmissions. We have
\begin{align}
\lim_{n\rightarrow\infty} \frac{k_j}{n} = R_j\frac{1-\frac{\prod_{k=1}^K\delta_k}{\delta_j}}{1-\prod_{k=1}^K\delta_k},
\end{align}
and thus
\begin{align}
\lim_{n\rightarrow\infty} \frac{k_j+k_j^{3/4}}{n} &= R_j\frac{1-\frac{\prod_{k=1}^K\delta_k}{\delta_j}}{1-\prod_{k=1}^K\delta_k},\\
\lim_{n\rightarrow\infty} \frac{n_1}{n}  &=  \max_{j\in\{1,\dots,K\}}\frac{R_j(1-\frac{\prod_{k=1}^K\delta_k}{\delta_j})}{(1-\delta_j)\frac{\prod_{k=1}^K\delta_k}{\delta_j}(1-\prod_{k=1}^K\delta_k)}.
\end{align}
Using Theorem~\ref{thm:1-to-k} the rate assertion of Theorem~\ref{thm:1-to-k_sec} follows.

\subsection{Dishonest user}
Similarly as in the honest-but-curious case, we need to compute $\lim_{n\rightarrow\infty}\frac{n_1}{n}$ and $\lim_{n\rightarrow\infty}\frac{\max\{n_2',n_2''\}}{n}$. It is immediate that
\begin{align}
\lim_{n\rightarrow\infty} \frac{n_2'}{n}&=\frac{R_1}{1-\delta_1}+\frac{R_2}{1-\delta_1\delta_2}\\
\lim_{n\rightarrow\infty} \frac{n_2''}{n}&=\frac{R_1}{1-\delta_1\delta_2}+\frac{R_2}{1-\delta_2}
\end{align}
Further 
\begin{align}
\lim_{n\rightarrow\infty} \frac{k_B}{n} &= R_1\frac{1-\delta_2}{1-\delta_1\delta_2}\\
\lim_{n\rightarrow\infty} \frac{k_C}{n} &= R_2\frac{1-\delta_1}{1-\delta_1\delta_2},
\end{align}
from which
\begin{align}
\lim_{n\rightarrow\infty} \frac{k_1}{n} &= R_1\frac{1-\delta_2}{\delta_2(1-\delta_1\delta_2)}\\
\lim_{n\rightarrow\infty} \frac{k_2}{n} &= R_2\frac{1-\delta_1}{\delta_1(1-\delta_1\delta_2)},
\end{align}
and 
\begin{align}
\lim_{n\rightarrow\infty} \frac{n_1}{n} &=\max\left( R_1\frac{1-\delta_2}{\delta_2(1-\delta_1)(1-\delta_1\delta_2)},R_2\frac{1-\delta_1}{\delta_1(1-\delta_2)(1-\delta_1\delta_2)}\right).
\end{align}
We also observe that
\begin{align}
R_1\frac{1-\delta_2}{\delta_2(1-\delta_1)(1-\delta_1\delta_2)}>R_2\frac{1-\delta_1}{\delta_1(1-\delta_2)(1-\delta_1\delta_2)} \Leftrightarrow\frac{R_1}{1-\delta_1}+\frac{R_2}{1-\delta_1\delta_2}>\frac{R_1}{1-\delta_1\delta_2}+\frac{R_2}{1-\delta_2}.
\end{align}
From these the rate assertion of Theorem~\ref{thm:dishonest} follows.

\section{Proof of distribution independent security}
\label{app:dis_sec_proof}
We need to show that if Bob is honest, then \eqref{eq:secrecy_dis_1} holds. In the proof we omit taking the maximum, but our argument is true for all joint distributions of $(W_1,W_2)$, hence the property follows.

We can almost identically follow the proof of Appendix~\ref{sec:analysis}. Similarly to \eqref{eq:security-analysis1} we have
\ifonecolumn
\begin{align}
I(W_1;Y_2^nS^n\Theta_2|W_2) &\leq I(W_1;Y_2^nS^n\Theta_2U_C|W_2)= I(W_1;Y_2^n|Y_2^{n_1+n_2}S^n\Theta_2U_CW_2). 
\end{align}
\else
\begin{align}
I(W_1;Y_2^nS^n\Theta_2|W_2) &\leq I(W_1;Y_2^nS^n\Theta_2U_C|W_2)
 \\&= I(W_1;Y_2^n|Y_2^{n_1+n_2}S^n\Theta_2U_CW_2). 
\end{align}
\fi
The last step follows because given $W_2$, variables $\Theta_A, \Theta_2, S^n$ are independent of $W_1$, further $Y_2^{n_1+n_2}, U^C$ are deterministic functions of $\Theta_A, \Theta_2, W_2, S^n$. 
The proofs of Lemmas~\ref{lem:secret key generation} and~\ref{lem:analysis:first-entropy-term} directly give us
\begin{align}
I(K_B;Y_2^{n_1+n_2}S^{n_1+n_2})&\leq k_Be^{-c_3\sqrt{k_1}},\\
H(Y_2^n|Y_2^{n_1+n_2}S^n\Theta_2U_CW_2) &\leq \Expect{M_B^C} ,
\end{align}
under the same conditions as defined in Lemmas~\ref{lem:secret key generation} and~\ref{lem:analysis:first-entropy-term}, where $c_3>0$ is some constant. We still need to show that
\ifonecolumn
\begin{align}
H(Y_2^n|W_1W_2Y_2^{n_1+n_2}S^n\Theta_2U_C)\geq
\Expect{\min\left(k_B,M_B^C\right)} -
I(K_B;Y_2^{n_1+n_2}S^{n_1+n_2})
\end{align}
\else
\begin{multline*}
H(Y_2^n|W_1W_2Y_2^{n_1+n_2}S^n\Theta_2U_C)\geq\\
\Expect{\min\left(k_B,M_B^C\right)} -
I(K_B;Y_2^{n_1+n_2}S^{n_1+n_2})
\end{multline*}
\fi
holds. We can again follow the proof of Lemma~\ref{lem:analysis:second-entropy-term}, but we have to argue the step
\ifonecolumn
\begin{align}
&I({K'}_B^C;W_1W_2Y_2^{n_1+n_2}S^n\Theta_2U_C|1_B^C) =I({K'}_B^C;Y_2^{n_1+n_2}S^{n_1+n_2}|1_B^C), \label{eq:to-be-argued}
\end{align}
\else
\begin{align}
&I({K'}_B^C;W_1W_2Y_2^{n_1+n_2}S^n\Theta_2U_C|1_B^C) \\&=I({K'}_B^C;Y_2^{n_1+n_2}S^{n_1+n_2}|1_B^C), \label{eq:to-be-argued}
\end{align}
\fi
where the independent and uniformly distributed property of $W_2$ was exploited when proving  the lemma. To see that equation \eqref{eq:to-be-argued} is true under the modified protocol, consider that $K'_C$ is generated from a different set of random packets than ${K'}_B^C$, so ${K'}_B^C - Y_2^{n_1+n_2} - U_C$ is Markov-chain, and since $(\Theta_A, S^n)$ is generated independently of $(W_1,W_2,\Theta_2)$, ${K'}_B^C - (Y_2^{n_1+n_2},1_B^C ,S^{n_1+n_2})- (W_1,W_2,\Theta_2,U_C)$ has the Markov property too.

Having established the three key lemmas for the modified protocol, we can conclude the proof the same way as we have seen in Section~\ref{sec:dishonest_scheme}. We omit the details to avoid repetitive arguments.

\section{Proofs of Lemmas in Section~\ref{sec:converse}}
We note that the order of proofs does not follow the order of appearance of the lemmas.
\label{app:converse_lemmas}
\subsection{Proof of Lemma~\ref{lemma:term_4}}
\label{app:term_4}
\begin{proof}
\ifonecolumn
\begin{align}
&n(R_1+R_2+R_3)-\E_3(1-\delta_1\delta_2\delta_3) \leq I(Y_1^nY_2^nY_3^nS^n;W_1W_2W_3)  \\
&=\sumin I(Y_{1,i}Y_{2,i}Y_{3,i}S_i;W_1W_2W_3|Y_1^{i-1}Y_2^{i-1}Y_3^{i-1}S^{i-1})  \\
&=\sumin I(Y_{1,i}Y_{2,i}Y_{3,i};W_1W_2W_3|Y_1^{i-1}Y_2^{i-1}Y_3^{i-1}S^{i-1}S_i) \\
&=\sumin \Pr\{S_i\neq\emptyset\} I(Y_{1,i}Y_{2,i}Y_{3,i};W_1W_2W_3|Y_1^{i-1}Y_2^{i-1}Y_3^{i-1}S^{i-1},S_i\neq \emptyset)  \\
&=\sumin I(X_i;W_1W_2W_3|Y_1^{i-1}Y_2^{i-1}Y_3^{i-1}S^{i-1})(1-\delta_1\delta_2\delta_3)
 \end{align}
\else
\begin{align}
&n(R_1+R_2+R_3)-\E_3(1-\delta_1\delta_2\delta_3)  \\
&\leq I(Y_1^nY_2^nY_3^nS^n;W_1W_2W_3)  \\
&=\sumin I(Y_{1,i}Y_{2,i}Y_{3,i}S_i;W_1W_2W_3|Y_1^{i-1}Y_2^{i-1}Y_3^{i-1}S^{i-1})  \\
&=\sumin I(Y_{1,i}Y_{2,i}Y_{3,i};W_1W_2W_3|Y_1^{i-1}Y_2^{i-1}Y_3^{i-1}S^{i-1}S_i) \\
&=\sumin \Pr\{S_i\neq\emptyset\}\cdot \\&\cdot I(Y_{1,i}Y_{2,i}Y_{3,i};W_1W_2W_3|Y_1^{i-1}Y_2^{i-1}Y_3^{i-1}S^{i-1},S_i\neq \emptyset)  \\
&=\sumin I(X_i;W_1W_2W_3|Y_1^{i-1}Y_2^{i-1}Y_3^{i-1}S^{i-1})(1-\delta_1\delta_2\delta_3)
 \end{align}
\fi
Here, the first inequality is Fano's inequality, besides, we exploited the independence property of $S_i$. This completes the proof of the first inequality of the lemma. Further, we also see that
\begin{align}
I(Y_1^nY_2^nY_3^nS^n;W_1W_2W_3)\leq n(R_1+R_2+R_3),
\end{align}
which by a similar argument gives the second inequality of the lemma.
\end{proof}

\subsection{Proof of Lemma~\ref{lemma:term_2}}
\label{app:term_2}
\begin{proof}
From the same type of derivation as we apply in Lemma~\ref{lemma:term_4}, we have that
\begin{align}
\sumin I(X_i;W_1|Y_1^{i-1}S^{i-1}) &\geq \frac{nR_1}{1-\delta_1}-\E_{1} \\
\sumin I(X_i;W_1|Y_1^{i-1}Y_2^{i-1}S^{i-1}) &\leq \frac{nR_1}{1-\delta_1\delta_2}.
\end{align}
Thus,
\ifonecolumn
\begin{align}
&\frac{nR_1}{1-\delta_1}-\E_1\leq\sumin
  I(X_i;W_1|Y_1^{i-1}S^{i-1}) \\ 
  &=\sumin I(X_i;W_1|Y_1^{i-1}Y_2^{i-1}S^{i-1}) +I(X_i;Y_2^{i-1}|Y_1^{i-1}S^{i-1})-I(X_i;Y_2^{i-1}|Y_1^{i-1}S^{i-1}W_1)\\
  &\leq\sumin I(X_i;W_1|Y_1^{i-1}Y_2^{i-1}S^{i-1}) +I(X_i;Y_2^{i-1}|Y_1^{i-1}S^{i-1}) \leq \frac{nR_1}{1-\delta_1\delta_2} + \sumin I(X_i;Y_2^{i-1}|Y_1^{i-1}S^{i-1}) \label{eq:lemma3_8}
 \end{align}
\else
\begin{align}
&\frac{nR_1}{1-\delta_1}-\E_1\leq\sumin
  I(X_i;W_1|Y_1^{i-1}S^{i-1}) \\ 
  &=\sumin I(X_i;W_1|Y_1^{i-1}Y_2^{i-1}S^{i-1})  \\
  &\quad+I(X_i;Y_2^{i-1}|Y_1^{i-1}S^{i-1})-I(X_i;Y_2^{i-1}|Y_1^{i-1}S^{i-1}W_1)\\
  &\leq\sumin I(X_i;W_1|Y_1^{i-1}Y_2^{i-1}S^{i-1})  \\
  &\quad+I(X_i;Y_2^{i-1}|Y_1^{i-1}S^{i-1}) \\
  &\leq \frac{nR_1}{1-\delta_1\delta_2} + \sumin I(X_i;Y_2^{i-1}|Y_1^{i-1}S^{i-1}) \label{eq:lemma3_8}
 \end{align}
 \fi
\end{proof}

\subsection{Proof of Lemma~\ref{lemma:term_3}}
\label{app:term_3}
\begin{proof}
The proof follows the same kind of derivation as the proof of Lemma~\ref{lemma:term_2}. We omit details to avoid repetition.
\end{proof}

\subsection{Proof of Lemma~\ref{lemma:balance}}
\label{app:balance}
\begin{proof}
We apply the shorthand $W^3=W_1W_2W_3$.
\ifonecolumn
\begin{align}
0&\leq{}H(Y_1^nS^n|Y_2^nY_3^nS^nW^3)={}H(Y_1^{n-1}S^{n-1}|Y_2^nY_3^nS^nW^3)+H(Y_{1,n}S_n|Y_1^{n-1}Y_2^nY_3^nS^nW^3) \\
&={}H(Y_1^{n-1}S^{n-1}|Y_2^{n-1}Y_3^{n-1}S^{n-1}W^3)-I(Y_1^{n-1}S^{n-1};Y_{2,n}Y_{3,n}S_n|Y_2^{n-1}Y_3^{n-1}S^{n-1}W^3)+ H(Y_{1,n}|Y_1^{n-1}Y_2^nY_3^nS^nW^3) \\ 
&={}H(Y_1^{n-1}S^{n-1}|Y_2^{n-1}Y_3^{n-1}S^{n-1}W^3)-I(Y_1^{n-1}S^{n-1};Y_{2,n}Y_{3,n}|Y_2^{n-1}Y_3^{n-1}S^{n-1}S_nW^3)+ H(Y_{1,n}|Y_1^{n-1}Y_2^nY_3^nS^nW^3)\\ 
&={}H(Y_1^{n-1}S^{n-1}|Y_2^{n-1}Y_3^{n-1}S^{n-1}W^3)-I(Y_1^{n-1}S^{n-1};Y_{2,n}Y_{3,n}|Y_2^{n-1}Y_3^{n-1}S^{n-1}S_n\notin\{\emptyset,\{1\}\} W^3)\Prob{S_n\notin\{\emptyset,\{1\}\}}\\
&\quad+ H(Y_{1,n}|Y_1^{n-1}Y_2^nY_3^nS^{n-1},S_n=\{1\},W^3)\Prob{S_n=\{1\}}\\
&={}H(Y_1^{n-1}S^{n-1}|Y_2^{n-1}Y_3^{n-1}S^{n-1}W^3)-I(Y_1^{n-1}S^{n-1};X_n|Y_2^{n-1}Y_3^{n-1}S^{n-1}W^3)(1-\delta_2\delta_3)\\
&\quad+ H(X_n|Y_1^{n-1}Y_2^{n-1}Y_3^{n-1}S^{n-1}W^3)(1-\delta_1)\delta_2\delta_3
\end{align}
\else
\begin{align}
&0\leq{}H(Y_1^nS^n|Y_2^nY_3^nS^nW^3)\\
&={}H(Y_1^{n-1}S^{n-1}|Y_2^nY_3^nS^nW^3)\\
&\quad+H(Y_{1,n}S_n|Y_1^{n-1}Y_2^nY_3^nS^nW^3) \\
&={}H(Y_1^{n-1}S^{n-1}|Y_2^{n-1}Y_3^{n-1}S^{n-1}W^3)\\
&\quad-I(Y_1^{n-1}S^{n-1};Y_{2,n}Y_{3,n}S_n|Y_2^{n-1}Y_3^{n-1}S^{n-1}W^3)\\
&\quad+ H(Y_{1,n}|Y_1^{n-1}Y_2^nY_3^nS^nW^3) \\ 
&={}H(Y_1^{n-1}S^{n-1}|Y_2^{n-1}Y_3^{n-1}S^{n-1}W^3)\\
&\quad-I(Y_1^{n-1}S^{n-1};Y_{2,n}Y_{3,n}|Y_2^{n-1}Y_3^{n-1}S^{n-1}S_nW^3) \\
&\quad+ H(Y_{1,n}|Y_1^{n-1}Y_2^nY_3^nS^nW^3)\\ 
&={}H(Y_1^{n-1}S^{n-1}|Y_2^{n-1}Y_3^{n-1}S^{n-1}W^3)\\
&\quad-I(Y_1^{n-1}S^{n-1};Y_{2,n}Y_{3,n}|Y_2^{n-1}Y_3^{n-1}S^{n-1}S_n\notin\{\emptyset,\{1\}\} W^3)\cdot \\
&\quad\quad\cdot\Prob{S_n\notin\{\emptyset,\{1\}\}}\\
&\quad+ H(Y_{1,n}|Y_1^{n-1}Y_2^nY_3^nS^{n-1},S_n=\{1\},W^3)\cdot\\
&\quad\quad\cdot\Prob{S_n=\{1\}}\\
&={}H(Y_1^{n-1}S^{n-1}|Y_2^{n-1}Y_3^{n-1}S^{n-1}W^3)\\
&\quad-I(Y_1^{n-1}S^{n-1};X_n|Y_2^{n-1}Y_3^{n-1}S^{n-1}W^3)(1-\delta_2\delta_3)\\
&\quad+ H(X_n|Y_1^{n-1}Y_2^{n-1}Y_3^{n-1}S^{n-1}W^3)(1-\delta_1)\delta_2\delta_3
\end{align}
\fi
 We do the same steps recursively to obtain the statement of the lemma.
\end{proof}

\subsection{Proof of Lemma~\ref{lemma:security}}
\label{app:secrecy}
\begin{proof}
From  (\ref{eq:secrecy}), we have 
\ifonecolumn
\begin{align}
&\epsilon>{} I(Y_2^nY_3^nS^n;W_1)={}\sumin I(X_i;W_1|Y_2^{i-1}Y_3^{i-1}S^{i-1})(1-\delta_2\delta_3)
 \end{align}
\else
\begin{align}
&\epsilon>{} I(Y_2^nY_3^nS^n;W_1)\\
%&=\sumin I(Y_{2,i}Y_{3,i}S_i;W_1|Y_2^{i-1}Y_3^{i-1}S^{i-1}) \\
%&={} \sumin I(Y_{2,i}Y_{3,i};W_1|Y_2^{i-1}Y_3^{i-1}S^{i-1}S_i)   \\
%&={} \sumin I(Y_{2,i}Y_{3,i};W_1|Y_2^{i-1}Y_3^{i-1}S^{i-1}S_i\notin\{\emptyset,B\})\cdot\\
%&\quad\quad\cdot\Pr\{S_i\notin\{\emptyset,B\}\}   \\
&={}\sumin I(X_i;W_1|Y_2^{i-1}Y_3^{i-1}S^{i-1})(1-\delta_2\delta_3)
 \end{align}
\fi
We omitted the intermediate steps that are in the same vein as in the proof of Lemma~\ref{lemma:term_4}.
\end{proof}

\subsection{Proof of Lemma~\ref{lemma:term_1_helper}}
\label{app:term_1_helper}
\begin{proof}
\ifonecolumn
\begin{multline}
\sumin I(X_i;Y_1^{i-1}|Y_2^{i-1}Y_3^{i-1}S^{i-1}W_1W_2W_3) =\\\sumin I(X_i;Y_1^{i-1}|Y_2^{i-1}Y_3^{i-1}S^{i-1}W_1)  - I(X_i;W_2W_3|Y_2^{i-1}Y_3^{i-1}S^{i-1}W_1)  + I(X_i;W_2W_3|Y_1^{i-1}Y_2^{i-1}Y_3^{i-1}S^{i-1}W_1)
\end{multline}
\else
\begin{align}
&\sumin I(X_i;Y_1^{i-1}|Y_2^{i-1}Y_3^{i-1}S^{i-1}W_1W_2W_3) \\
&=\sumin I(X_i;Y_1^{i-1}|Y_2^{i-1}Y_3^{i-1}S^{i-1}W_1) \\
&\quad - I(X_i;W_2W_3|Y_2^{i-1}Y_3^{i-1}S^{i-1}W_1) \\
&\quad + I(X_i;W_2W_3|Y_1^{i-1}Y_2^{i-1}Y_3^{i-1}S^{i-1}W_1)
\end{align}
\fi
From~(\ref{eq:decodability}) and Fano's inequality we have
\ifonecolumn
\begin{align}
I(Y_2^nY_3^nS^n;W_2W_3|W_1) \leq  I(Y_2^nY_3^nS^n;W_2W_3) - (h_2(\epsilon)+\epsilon ). 
\end{align}
\else
\begin{multline}
I(Y_2^nY_3^nS^n;W_2W_3|W_1) \leq \\ I(Y_2^nY_3^nS^n;W_2W_3) - (h_2(\epsilon)+\epsilon ). 
\end{multline}
\fi
We expand these terms the same way as we did in the proof of Lemma~\ref{lemma:term_4}, and we can write 
\ifonecolumn
\begin{align}
\sumin I(X_i;W_2W_3|Y_2^{i-1}Y_3^{i-1}S^{i-1}W_1) \leq \sumin I(X_i;W_2W_3|Y_2^{i-1}Y_3^{i-1}S^{i-1})+\E_{5}. 
\end{align}
\else
\begin{multline}
\sumin I(X_i;W_2W_3|Y_2^{i-1}Y_3^{i-1}S^{i-1}W_1) \leq \\\sumin I(X_i;W_2W_3|Y_2^{i-1}Y_3^{i-1}S^{i-1})+\E_{5}, 
\end{multline}
\fi
From the independence property of the messages
\ifonecolumn
\begin{align}
\sumin I(X_i;W_2W_3|Y_1^{i-1}Y_2^{i-1}Y_3^{i-1}S^{i-1}W_1) \geq \sumin I(X_i;W_2W_3|Y_1^{i-1}Y_2^{i-1}Y_3^{i-1}S^{i-1}), 
\end{align}
\else
\begin{multline}
\sumin I(X_i;W_2W_3|Y_1^{i-1}Y_2^{i-1}Y_3^{i-1}S^{i-1}W_1) \\\geq \sumin I(X_i;W_2W_3|Y_1^{i-1}Y_2^{i-1}Y_3^{i-1}S^{i-1}), 
\end{multline}
\fi
where $\E_{5}=\frac{h_2(\epsilon)+\epsilon }{1-\delta_2\delta_3}$.
This enables us to use the same idea as in Lemma~\ref{lemma:term_4} to bound these terms. Doing so gives us
\ifonecolumn
\begin{align}
&\sumin I(X_i;Y_1^{i-1}|Y_2^{i-1}Y_3^{i-1}S^{i-1}W_1W_2W_3) \geq\sumin I(X_i;Y_1^{i-1}|Y_2^{i-1}Y_3^{i-1}S^{i-1}W_1) \label{eq:to_bound}- \frac{n(R_2+R_3)}{1-\delta_2\delta_3} +\frac{n(R_2+R_3)}{1-\delta_1\delta_2\delta_3} -\E_{2}'-\E_5, 
\end{align}
\else
\begin{align}
&\sumin I(X_i;Y_1^{i-1}|Y_2^{i-1}Y_3^{i-1}S^{i-1}W_1W_2W_3) \\
&\geq\sumin I(X_i;Y_1^{i-1}|Y_2^{i-1}Y_3^{i-1}S^{i-1}W_1) \label{eq:to_bound}\\
&\quad - \frac{n(R_2+R_3)}{1-\delta_2\delta_3} +\frac{n(R_2+R_3)}{1-\delta_1\delta_2\delta_3} -\E_{2}'-\E_5, 
\end{align}
\fi
where $\E_2'=\frac{h_2(2\epsilon)+2\epsilon }{1-\delta_2\delta_3}$.
It remains to give a bound on term (\ref{eq:to_bound}). From Lemma~\ref{lemma:security} and after a few basic steps we can arrive to
\ifonecolumn
\begin{align}
\E_4 &> \sumin I(X_i;W_1|Y_2^{i-1}Y_3^{i-1}S^{i-1}) \\&=\sumin
  -I(X_i;Y_1^{i-1}|Y_2^{i-1}Y_3^{i-1}S^{i-1}W_1)
  +I(X_i;W_1|Y_1^{i-1}Y_2^{i-1}Y_3^{i-1}S^{i-1})+ I(X_1;Y_1^{i-1}|Y_2^{i-1}Y_3^{i-1}S^{i-1}) \label{eq:back1}.
\end{align}
\else
\begin{align}
&\E_4 > \sumin I(X_i;W_1|Y_2^{i-1}Y_3^{i-1}S^{i-1}) \\
&=\sumin
  -I(X_i;Y_1^{i-1}|Y_2^{i-1}Y_3^{i-1}S^{i-1}W_1)\\
  &\quad+I(X_i;W_1|Y_1^{i-1}Y_2^{i-1}Y_3^{i-1}S^{i-1})\\
  &\quad+ I(X_1;Y_1^{i-1}|Y_2^{i-1}Y_3^{i-1}S^{i-1}) \label{eq:back1}.
\end{align}
\fi
From a similar result as in Lemma~\ref{lemma:term_4}:
\begin{align}
I(X_i;W_1|Y_1^{i-1}Y_2^{i-1}Y_3^{i-1}S^{i-1}) \geq \frac{nR_1}{1-\delta_1\delta_2\delta_3} - \E_{6},
\end{align}
where $\E_{6}=\frac{h_2(\epsilon)+\epsilon }{1-\delta_1\delta_2\delta_3}$.
Further, a symmetric result to Lemma~\ref{lemma:term_3} shows:
\ifonecolumn
\begin{align}
I(X_1;Y_1^{i-1}|Y_2^{i-1}Y_3^{i-1}S^{i-1})\frac{n(R_2+R_3)}{1-\delta_2\delta_3}-\frac{n(R_2+R_3)}{1-\delta_1\delta_2\delta_3} -\E_{2}'.
\end{align}
\else
\begin{multline*}
I(X_1;Y_1^{i-1}|Y_2^{i-1}Y_3^{i-1}S^{i-1})\geq \\\frac{n(R_2+R_3)}{1-\delta_2\delta_3}-\frac{n(R_2+R_3)}{1-\delta_1\delta_2\delta_3} -\E_{2}'.
\end{multline*}
\fi
Applying these bounds in (\ref{eq:back1}) and then substituting back to (\ref{eq:to_bound}) results the claim of the lemma.
\end{proof}

\section{Proof of Lemma~\ref{lem:equivalence}}
\label{app:equivalence}
\begin{proof}
As a first step we define
\ifonecolumn
\begin{align}
{\mathbf{Adv}}^{*\text{ss}}_{\text{dis}} = \max_{f,P_{W_1},w_2,\s} \left \{ \max_{\mathcal{A}} \Prob{\A(Y_2^n,S^n,\Theta_2,\s,w_2)=f(W_1,w_2)}-\max_\S \Prob{\S(P_{W_1},f,w_2)=f(W_1,w_2)}\right\}.
\end{align}
\else
\begin{multline*}
{\mathbf{Adv}}^{*\text{ss}}_{\text{dis}} =\\ \max_{f,P_{W_1},w_2,\s} \left \{ \max_{\mathcal{A}} \Prob{\A(Y_2^n,S^n,\Theta_2,\s,w_2)=f(W_1,w_2)}\right. \\\left.-\max_\S \Prob{\S(P_{W_1},f,w_2)=f(W_1,w_2)}\right\}.
\end{multline*}
\fi
As opposed to \Advss\ here $w_2$ is not a random variable but a constant value from $\mathcal{W}_2$. Clearly,
\begin{align}
\Advssdiss\leq\Advssdis,
\end{align}
because $W_2$ taking the value $w_2$ with probability~1 is a particular joint distribution $W_1,W_2$ can take, so the scope of the maximization is restricted. We show that $\Advssdiss=\Advssdis$.
\ifonecolumn
\begin{align}
{\mathbf{Adv}}^{\text{ss}}_{\text{dis}} &= \max_{f,P_{W_1,W_2},\s} \left \{ \max_{\mathcal{A}} \Prob{\A(Y_2^n,S^n,\Theta_2,\s,W_2)=f(W_1,W_2)}-\max_\S \Prob{\S(P_{W_1},f,W_2)=f(W_1,W_2)}\right\} \\
%&=\max_{P_{W_2}} \max_{f,P_{W_1|W_2}}\left \{ \max_{\mathcal{A}} \Prob{\A(Y_2^nS^n\Theta_2,W_2)=f(W_1,W_2)} -\max_\S \Prob{\S(P_{W_1},f,W_2)=f(W_1,W_2)}\right\} \\
&=\max_{f,P_{W_1,W_2},\s} \sum_{w_2} p_{W_2}(w_2)\left\{ \max_{\mathcal{A}} \Prob{\A(Y_2^n,S^n,\Theta_2,\s,W_2)=f(W_1,W_2)|W_2=w_2} \right.\\
&\quad\left.\quad\quad-\max_\S \Prob{\S(P_{W_1},f,\s,W_2)=f(W_1,W_2)|W_2=w_2}\right\} \\
&=\max_{w_2^*,f,P_{W_1|W_2=w_2^*},\s} \left \{ \max_{\mathcal{A}} \Pr\{\A(Y_2^nsS^ns\Theta_2,\s,w_2^*)=f(W_1,w_2^*)\} -\max_\S \Prob{\S(P_{W_1},f,w_2^*)=f(W_1,w_2^*)}\right\} \\
&=\max_{f,P_{W_1,w_2^*},\s} \left \{ \max_{\mathcal{A}} \Prob{\A(Y_2^n,S^n,\Theta_2,\s,w_2^*)=f(W_1,w_2^*)}-\max_\S \Prob{\S(P_{W_1},f,w_2^*)=f(W_1,w_2^*)}\right\} \\
&=\Advssdiss.
\end{align}
\else
\begin{align}
&{\mathbf{Adv}}^{\text{ss}}_{\text{dis}} = \\&\max_{f,P_{W_1,W_2},\s} \left \{ \max_{\mathcal{A}} \Prob{\A(Y_2^n,S^n,\Theta_2,\s,W_2)=f(W_1,W_2)}\right.\\&\left. -\max_\S \Prob{\S(P_{W_1},f,W_2)=f(W_1,W_2)}\right\} \\
%&=\max_{P_{W_2}} \max_{f,P_{W_1|W_2}}\left \{ \max_{\mathcal{A}} \Prob{\A(Y_2^nS^n\Theta_2,W_2)=f(W_1,W_2)} -\max_\S \Prob{\S(P_{W_1},f,W_2)=f(W_1,W_2)}\right\} \\
&=\max_{f,P_{W_1,W_2},\s} \sum_{w_2} p_{W_2}(w_2)\cdot\\&\quad \cdot\left\{ \max_{\mathcal{A}} \Prob{\A(Y_2^n,S^n,\Theta_2,\s,W_2)=f(W_1,W_2)|W_2=w_2} \right.\\
&\quad\left.-\max_\S \Prob{\S(P_{W_1},f,\s,W_2)=f(W_1,W_2)|W_2=w_2}\right\} \\
&=\max_{w_2^*,f,P_{W_1|W_2=w_2^*},\s} \left \{ \max_{\mathcal{A}} \Pr\{\A(Y_2^nsS^ns\Theta_2,\s,w_2^*)=\right.\\&\quad\left.f(W_1,w_2^*)\} -\max_\S \Prob{\S(P_{W_1},f,w_2^*)=f(W_1,w_2^*)}\right\} \\
&=\max_{f,P_{W_1,w_2^*},\s} \left \{ \max_{\mathcal{A}} \Prob{\A(Y_2^n,S^n,\Theta_2,\s,w_2^*)=f(W_1,w_2^*)} \right.\\&\quad\left.-\max_\S \Prob{\S(P_{W_1},f,w_2^*)=f(W_1,w_2^*)}\right\} \\
&=\Advssdiss.
\end{align}
\fi
where the second step follows because there is a certain value $w_2^*$ of $W_2$ that maximizes the expression inside $\{\dots\}$, and moreover this expression depends on $P_{W_1|W_2}$ only through $P_{W_1|W_2=w_2}$, hence a maximizing joint distribution of $W_1,W_2$ is when $W_2$ takes this particular value with probability~1.

We continue the proof in two steps, first we define a notion of distinguishing security applicable for jointly distributed messages by extending a similar definition in \cite{Vardy12} and show its equivalence with the above definition of semantic security. Then we show equivalence between this notion of distinguishing security and distribution independent security as defined by~\Advmisdis.

We define a notion corresponding to distinguishing security by defining the adversarial advantage:
\ifonecolumn
\begin{align}
\Advdsdis = \max_{\A,w_1^0,w_1^1,w_2,\s} 2\Prob{\A(w_1^0,w_1^1,w_2,{}^bY_2^n,S^n,\Theta_2,\s)=b}-1,
\end{align}
\else
\begin{multline*}
\Advdsdis = \\\max_{\A,w_1^0,w_1^1,w_2,\s} 2\Prob{\A(w_1^0,w_1^1,w_2,{}^bY_2^n,S^n,\Theta_2,\s)=b}-1,
\end{multline*}
\fi
where $w_1^0,w_1^1 \in \mathcal{W}_1$ are possible messages, similarly $w_2 \in \mathcal{W}_2$, $b$ is a variable uniformly distributed over $\{0,1\}$ and is independent of all other variables, while ${}^bY_2^n$ is Calvin's observation given $W_1=w_1^b$.

Distinguishing security defined by \Advdsdis\ is equivalent to semantic security as defined by \Advssdiss\ and  hence equivalently as defined by \Advssdis. To show that distinguishing security implies semantic security, we can almost identically follow the proof of Theorem~5 from \cite{Vardy12}, with a slight difference that a conditioning on $W_2$ appears. 
%%%%%%%%%%%    These are details, maybe can be omitted %%%%%%%%%%%%%%%
Given an adversary $\A_{ss}$ attacking semantic security, we construct an adversary $\A_{ds}$ attacking distinguishing security as follows: $\A_{ds}$ outputs~1, if the adversary attacking semantic security $\A_{ss}$ gives as output $f(w_1^1,w_2)$, otherwise it returns 0. Then, if $W_1^0$ and $W_1^1$ are i.i.d.~both having the same distribution as $W_1$, then
\ifonecolumn
\begin{align}
&\Prob{\A_{ds}(W_1^0,W_1^1, W_2,{}^1Y_2^n,\S^n,\Theta_2,\s)=1|W_2=w_2}=\Prob{A_{ss}(Y_2^n,S^n,\Theta_2,\s,W_2)=f(W_1,W_2)|W_2=w_2}\\
&\Prob{\A_{ds}(W_1^0,W_1^1, W_2,{}^0Y_2^n,S^n,\Theta_2,\s)=1|W_2=w_2}\leq \max_\S \Prob{\S(P_{W_1},f,W_2)=f(W_1,W_2)|W_2=w_2}.
\end{align}
\else
\begin{align}
&\Prob{\A_{ds}(W_1^0,W_1^1, W_2,{}^1Y_2^n,\S^n,\Theta_2,\s)=1|W_2=w_2}\\&=\Prob{A_{ss}(Y_2^n,S^n,\Theta_2,\s,W_2)=f(W_1,W_2)|W_2=w_2}\\
&\Prob{\A_{ds}(W_1^0,W_1^1, W_2,{}^0Y_2^n,S^n,\Theta_2,\s)=1|W_2=w_2}\\&\leq \max_\S \Prob{\S(P_{W_1},f,W_2)=f(W_1,W_2)|W_2=w_2}.
\end{align}
\fi
Finishing the derivation as in \cite{Vardy12} we get
\ifonecolumn
\begin{multline*}
\Prob{A_{ss}(Y_2^n,S^n,\Theta_2,\s,W_2)=f(W_1,W_2)|W_2=w_2} - \max_\S \Prob{\S(P_{W_1},f,W_2)=f(W_1,W_2)|W_2=w_2}\\\leq
 \max_{w_1^0,w_1^1,w_2,\A_{ds},\s} 2\Prob{\A_{ds}(w_1^0,w_1^1, w_2,{}^bY_2^n,S^n,\Theta_2,\s)=b}-1
\end{multline*}
\else
\begin{align}
&\Prob{A_{ss}(Y_2^n,S^n,\Theta_2,\s,W_2)=f(W_1,W_2)|W_2=w_2} \\&\quad- \max_\S \Prob{\S(P_{W_1},f,W_2)=f(W_1,W_2)|W_2=w_2}\leq\\
& \max_{w_1^0,w_1^1,w_2,\A_{ds},\s} 2\Prob{\A_{ds}(w_1^0,w_1^1, w_2,{}^bY_2^n,S^n,\Theta_2,\s)=b}-1
\end{align}
\fi
for all $P_{W_1},f,\A_{ss},\s$, hence taking the maximum over these variables on the LHS and over $w_2$ on both sides keeps the inequality.
%%%%%%%%%%%%%%%%%%%%%%%%%%%%%%%%%%%%%%%%%%%%%%%
This establishes that
\begin{align}
\Advssdis =\Advssdiss \leq \Advdsdis \leq 2\Advssdis.
\end{align}
The other direction of implication is a straightforward consequence of the definitions, the scope of maximization in \Advdsdis\ is a subset of that of \Advssdis, in case of \Advdsdis\ $f$ is a function that computes $b$, while $P_{W_1,W_2}$ is such that $W_1$ uniformly takes the two values $w_1^0$ and $w_1^1$ and independently $W_2$ takes $w_2$ with probability~1.

What remains to show is that distinguishing security defined by \Advdsdis\ is equivalent to distribution independent security as defined by \Advmisdis.
Clearly, for any particular value of $w_2$,
\begin{align}
\Advmisdis \geq \max_{P_{W_1},\s} I(W_1; Y_2^nS^n\Theta_2|W_2=w_2).
\end{align}
If we fix $w_2$ for the scheme, we can directly invoke Theorem~5 from \cite{Vardy12} which proves that
\ifonecolumn
\begin{align}
\max_{\A,w_1^0,w_1^1,\s} 2\Prob{\A(w_1^0,w_1^1,w_2,{}^bY_2^n,S^n,\Theta_2,\s)=b}-1\leq \sqrt{2\max_{P_{W_1},\s} I(W_1; Y_2^nS^n\Theta_2|W_2=w_2)}\leq \sqrt{2\cdot\Advmisdis},
\end{align}
\else
\begin{multline*}
\max_{\A,w_1^0,w_1^1,\s} 2\Prob{\A(w_1^0,w_1^1,w_2,{}^bY_2^n,S^n,\Theta_2,\s)=b}-1\leq\\ \sqrt{2\max_{P_{W_1},\s} I(W_1; Y_2^nS^n\Theta_2|W_2=w_2)}\leq \sqrt{2\cdot\Advmisdis},
\end{multline*}
\fi
which holds for every $w_2$, so we can take the maximum in $w_2$ on the LHS, which gives in turn
\begin{align}
\Advdsdis \leq \sqrt{2\cdot\Advmisdis}
\end{align}
showing that the distribution independent security implies distinguishing security. The other direction is also true. We can apply the same type of argument as when showing $\Advssdis=\Advssdiss$ to get:
\begin{align}
{\mathbf{Adv}}^{\text{mis}}_{\text{dis}}&= \max_{P_{W_1,W_2},\s}I(W_1;Y_2^nS^n\Theta_2|W_2)\\& = \max_{w_2,P_{W_1},\s}I(W_1;Y_2^nS^n\Theta_2|W_2=w_2).
\end{align}
Let us denote 
\ifonecolumn
\begin{align}
\Advds(w_2) =  \max_{\A,w_1^0,w_1^1,\s} 2\Prob{\A(w_1^0,w_1^1,w_2,{}^bY_2^n,S^n,\Theta_2,\s)=b}-1,
\end{align}
\else
\begin{multline*}
\Advds(w_2) =\\  \max_{\A,w_1^0,w_1^1,\s} 2\Prob{\A(w_1^0,w_1^1,w_2,{}^bY_2^n,S^n,\Theta_2,\s)=b}-1,
\end{multline*}
\fi
We can apply Theorem~4.9 from \cite{Vardy12} with a conditioning on $W_2=w_2$, which implies that for any $w_2$:
\ifonecolumn
\begin{align}
\max_{P_{W_1},\s}I(W_1;Y_2^nS^n\Theta_2|W_2=w_2)\leq  2\Advds(w_2)\log\left( \frac{2^n}{\Advds(w_2)}  \right).
\end{align}
\else
\begin{multline*}
\max_{P_{W_1},\s}I(W_1;Y_2^nS^n\Theta_2|W_2=w_2)\leq \\ 2\Advds(w_2)\log\left( \frac{2^n}{\Advds(w_2)}  \right).
\end{multline*}
\fi
Since the above is true for any $w_2$, we can take the maximum in $w_2$ on both sides resulting
\begin{align}
\Advmisdis \leq  2\Advds\log\left( \frac{2^n}{\Advds}  \right).
\end{align}
This completes the proof that distribution independent security is equivalent to semantic security defined by $\Advssdis$.
\end{proof}
\end{document}